\documentclass[preprint,5p,times,authoryear]{elsarticle}

\usepackage{amssymb}
\usepackage{amsmath}    

\usepackage[singlelinecheck=false]{caption}

 \usepackage{url}

\usepackage{cleveref}   

\usepackage{supertabular}   

\graphicspath{{figures/}}
\usepackage[usenames, dvipsnames]{color}    
\usepackage[normalem]{ulem}

\newcommand{\bqn}{\begin{eqnarray}}
\newcommand{\eqn}{\end{eqnarray}}
\newcommand{\bq}{\begin{eqnarray*}}
\newcommand{\eq}{\end{eqnarray*}}

\journal{arXiv}

\begin{document}

\begin{frontmatter}

\title{Reviews: Topological Distances and Losses for Brain Networks}

\author[uwm,wais]{Moo K. Chung\corref{cor}}
\ead{mkchung@wisc.edu}
\author[chem]{Alexander Smith}
\author[uwpd]{Gary Shiu}
\cortext[cor]{Correspondence to: M.K.Chung, Department of Biostatistics and Medical Informatics, University of Wisconsin-Madison, 1300 University Ave, 4725 Medical Science Center, Madison, 53706, USA}
\address[uwm]{Department of Biostatistics and Medical Informatics, University of Wisconsin, Madison, WI, USA}
\address[wais]{Waisman Laboratory for Brain Imaging and Behavior, University of Wisconsin, Madison, WI, USA}
\address[chem]{Department of Chemical and Biological Engineering, University of Wisconsin, Madison, WI, USA}
\address[uwpd]{Department of Physics, University of Wisconsin, Madison, WI, USA}

\begin{abstract}
Almost all statistical and machine learning methods in analyzing brain networks rely on distances and loss functions, which are mostly Euclidean or matrix norms. The Euclidean or matrix distances may fail to capture underlying subtle topological differences in brain networks. Further, Euclidean distances are sensitive to outliers. A few extreme edge weights may severely affect the distance. Thus it is necessary to use distances and loss functions  that recognize topology of data. In this review paper, we survey various topological distance and loss functions from topological data analysis (TDA) and persistent homology that can be used in brain network analysis more effectively. Although there are many recent brain imaging studies that are based on TDA methods, possibly due to the lack of method awareness, TDA has not taken as the mainstream tool in brain imaging field yet. The main purpose of this paper is provide the relevant technical survey of these powerful tools that are immediately applicable to brain network data.  
\end{abstract}

\begin{keyword}
Topological distances \sep topological losses \sep topological data analysis \sep persistent homology \sep brain networks  \end{keyword}

\end{frontmatter}

\section{Introduction}

There are many similarity measures, distances and loss functions used in discriminating  brain networks  \citep{banks.1994,chung.2019.CNI,chung.2019.NN,lee.2012.TMI,chen.2016}. Due to ever increasing popularity of deep learning, which estimates the parameters by minimizing loss functions, there is also a renewed interest in studying various loss functions. Many of existing distances and losses simply ignore the underlying topology of the networks and often use Euclidean distances. Existing distances often fail to capture underlying topological differences. They may lose sensitivity over topological structures such as the connected components, modules and cycles in networks. Further, Euclidean norms are sensitive to outliers. A few extreme edge weights may severely affect the distance.

In standard graph theory based brain network analysis,  the similarity of networks are measured by the difference in graph theory features such as assortativity, betweenness centrality, small-worldness and network homogeneity \citep{bullmore.2009,rubinov.2010.NI,uddin.2008}. Comparison of graph theory features appears to reveal changes of structural or functional connectivity associated with different clinical populations \citep{rubinov.2010.NI}. Since dense weighted brain networks are difficult to interpret and analyze, they are often turned into binary networks by thresholding edge weights \citep{he.2008,vanwijk.2010}. The choice of thresholding the edge weights may alter  the network topology. The multiple comparison correction over every possible connections were used in determining thresholds \citep{rubinov.2009.HBM,salvador.2005,vanwijk.2010}. The amount of sparsity of connections were also  used in determining thresholds \citep{bassett.2006,he.2008,vanwijk.2010,lee.2011.TMI}. To address the inherent problem of thresholding, persistent homological brain network analysis was developed in 2011 \citep{lee.2011.MICCAI,lee.2011.ISBI}. Since then the method has been successfully applied to numerous brain network studies  and shown to be a powerful alternative brain network approaches \citep{chung.2013.MICCAI,
chung.2015.TMI,lee.2012.TMI}.

Persistent homology provides a coherent mathematical framework for quantifying brain images and networks \citep{carlsson.2008,edelsbrunner.2010}. Instead of looking at network at a fixed thresholding or resolution, persistent homology quantifies the over all changes of network topology over multiple scales \citep{edelsbrunner.2010,horak.2009,zomorodian.2005}. In doing so, it reveals the most {\em persistent} topological features that are robust to noise and scale.  Persistent homology has been applied to wide variety of data including sensor networks \citep{carlsson.2008}, protein structures \citep{gameiro.2015} and RNA viruses \citep{chan.2013}, image segmentation.

In persistent homology based brain network analysis,  instead of analyzing networks at one fixed threshold that may not be optimal, we build the collection of nested networks over every possible threshold using the {\em graph filtration}, a persistent homological construct  \citep{lee.2011.MICCAI, lee.2012.TMI, chung.2013.MICCAI, chung.2015.TMI}. The graph filtration is a threshold-free framework for analyzing a family of graphs but requires hierarchically building specific nested subgraph structures. The graph filtration shares similarities to the existing multi-thresholding or multi-resolution network models  that use many different arbitrary thresholds or scales  \citep{achard.2006,he.2008,lee.2012.TMI,kim.2015,supekar.2008}. 
 Such approaches are mainly used to visually display the dynamic pattern of how graph theoretic features change over different thresholds and the pattern of change is rarely quantified. Persistent homology can be used to quantify such dynamic pattern in a more coherent mathematical framework. In numerous studies, persistent homological network approach is shown to very robust and outperforming many existing network measures and methods. In \citet{lee.2011.MICCAI,lee.2012.TMI}, persistent homology was shown to outperform against  eight existing graph theory features such as  assortativity, between centrality, clustering coefficient, characteristic path length, samll-worldness, modularity and global network homogeneity.  In \citet{chung.2017.CNI}, persistent homology was shown to outperform  various matrix norms. In \citet{wang.2018.annals}, persistent homology using the persistent landscape was shown to outperform power spectral density and local variance methods. In \citet{wang.2017.CNI}, persistent homology was shown to outperform topographic power maps. In \citet{yoo.2017}, center persistency was shown to outperform the network-based statistic and element-wise multiple corrections.

Existing statistical and machine learning methods for brain networks are based on distance and loss functions that are mostly Euclidean based or matrix norms. Such distances are geometric in nature and not sensitive enough for topological signal differences often observed in brain networks. Thus, it is necessary to use distances and loss functions  that are topologically more sensitive. Persistent homology offers a coherent mathematical framework for measuring network distances {\em topologically}. Various topological distances such as Gromov-Hausdorff (GH) distance\citep{tuzhilin.2016,carlsson.2008,carlsson.2010,chazal.2009,lee.2011.MICCAI,lee.2012.TMI}, bottleneck distances 
\citep{lee.2012.TMI,lee.2017.HBM,chung.2015.TMI} and the Kolmogorov-Smirnov (KS) distance\citep{chung.2012.CNA,chung.2017.IPMI,
lee.2017.HBM} are available. The main purpose of this paper is to review such topological distances within the context of persistent homology.

\section{Traditional distances and losses}

\subsection{Losses in statistics and machine learning}

In statistical analysis and machine learning, the loss function acts as a cost function, which needs to be minimized to determine the  model fit. Widely used loss functions are often Euclidean, and have proven effective in traditional applications \citep{bishop2006pattern}. Here, we describe the most often used loss functions that were applied  in various brain image analysis tasks.

The mean square error loss (MSE) is the most often used loss. For a given model with $n$ input data $x_i$, input labels $y_i$, and model outputs $\widehat{y}_i$, MSE is given as
\begin{equation}
MSE = \frac{1}{n} \sum_{i=1}^n (y_i - \widehat{y}_i)^2
\label{eq:mse}
\end{equation}
The MSE  is based upon the $L_2$-norm between the model prediction and input data and the most often used loss in regression. 
In regression setting, MSE was in identifying brain imaging predictions for memory performance \citep{wang2011sparse}, stimation of kinetic constants from PET data \citep{o1999use}, for correction partial volume effects in arterial spin labeling MRI \citep{kim2018improving,asllani2008regression}, EEG signal classification with neural networks \citep{kottaimalai2013eeg}. 
This is also the basis of widely used $k$-means clustering \citep{jain1999data} in identifying abnormalities in brain tumors \citep{arunkumar2019k}, modeling state spaces in rsfMRI \citep{huang.2020.NM}. 
This loss was also used for image classification of autism spectrum disorder \citep{heinsfeld2018identification}, tumor segmentation for MR brain images \cite{mittal2019deep} among other learning tasks.

One problem associated with MSE  is that it heavily penalizes outliers due to its quadratic lose \citep{bishop2006pattern}. The  impact of outliers can be significantly reduced using  the mean absolute error (MAE), which is $L_2$-norm:
\begin{equation}
MAE = \frac{1}{n} \sum_{i=1}^n |y_i - \widehat{y}_i|
\label{eq:mae}
\end{equation}
The MAE was successfully used in clustering \citep{jain1999data}, tumor segmentation in MRI \citep{blessy2015performance}, age prediction in deep learning \citep{cole2017predicting} and the conversion of MRI to CT through deep learning \citep{wolterink2017deep}.

Another common loss often used in probabilistic model building is the cross entropy $L_{ent}$, which was initially used in logistic regression and artificial neural networks. The output of the model is 
defined as a probability $p_i$ that a given input $x_i$ belongs to a binary class:
\begin{equation}
L_{ent} =  y_i \log p_i  + (1-y_i) \log (1- p_i).
\label{eq:lent}
\end{equation}
The cross entropy is equivalent to the maximum likelihood of the product of Bernoulli distributions. The use of cross entropy over MSE can lead to faster training times and improved model generalization \citep{bishop2006pattern}. The cross entropy has been used in improving brain MRI segmentation \citep{moeskops2017adversarial}, the prediction of intracranial pressure levels after brain injury \citep{chen2010intracranial} and in a machine learning interface for medical image analysis \citep{zhang2017machine}. Logistic regression has found use in a deep learning ensemble regression model for brain disease diagnosis \citep{suk2017deep} and in the classification of brain hemorrhages with the convolutional neural networks \citep{jnawali2018deep}.

Another broadly applied loss in the support vector machine and other classification tasks is the {\em hinge loss}, which uses the $L_{\infty}$-norm \citep{bishop2006pattern}. The hinge loss heavily penalizes incorrect classifications or classifications that are correct but are near the decision boundary. The hinge loss has been used in the classification of CT brain images with deep learning networks  for identification of Alzheimer's disease \citep{gao2017classification}. It has also been applied to data collected from working brains to guide machine learning algorithms \citep{fong2018using}.

\subsection{Matrix norms as network distances}

Many distance or similarity measures are not metrics but having metric distances makes networks more stable due to the triangle inequality.  Further, existing network distance concepts are often borrowed from  the metric space theory. Let us start with formulating brain networks as metric spaces. The brain networks are often algebraically represented as matrices of size $p \times p$, where $p$ is predetermined number of parcellations. Often 100-300 parcellations are used for this purpose. It is necessary to use distances or losses defined on the connectivity matrices. Consider a weighted graph or network $\mathcal{X} = (V, w)$ with the node set $V = \left\{ 1, \dots, p \right\}$ and the edge weights $w=(w_{ij})$, where $w_{ij}$ is the weight between nodes $i$ and $j$.
The edge weight is usually given by a similarity measure between the observed data on the nodes. Various similarity measures  have been proposed. The correlation or mutual information between measurements for the biological or metabolic network and the frequency of contact between actors for the social network have been used as edge weights \citep{bassett.2006,bien.2011}.
We may assume that the edge weights satisfy the metric properties: nonnegativity, identity, symmetry and the triangle inequality such that
$$w_{i,j} \geq 0, \; w_{ii} =0,\; w_{ij} = w_{ji}, \;w_{ij} \leq w_{ik} + w_{kj}.$$ 
With theses conditions, $\mathcal{X}=(V, w)$ forms a metric space. Although the metric property is not necessary for building a network, it offers many nice mathematical properties and easier interpretation on network connectivity. Further, persistent homology is often built on top of metric spaces. Many real-world networks satisfy the metric properties.

Given measurement vector ${\bf x}_i  = (x_{1i}, \cdots, x_{ni})^{\top} \in \mathbb{R}^n$ on the node $i$. 
The weight $w=(w_{ij})$ between nodes is often given by some bivariate function $f$: $w_{ij} = f({\bf x}_i, {\bf x}_j)$. The correlation between ${\bf x}_i$ and ${\bf x}_j$, denoted as $\mbox{corr}({\bf x}_i, {\bf x}_j)$, is a bivariate function. If the weights $w=(w_{ij})$ are given by 
$$w_{ij} = \sqrt{1-\mbox{corr}({\bf x}_i, {\bf x}_j)},$$  
it can be shown that $\mathcal{X}=(V, w)$ forms a metric space.

\begin{figure}
\centering
\includegraphics[width=0.7\linewidth]{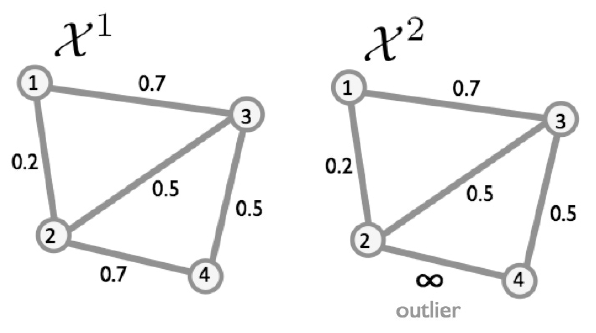}
\caption{Toy networks with an outlying edge weight. All the $L_l$-norm based distance will give $\infty$ distance while topological distances may not be severely affected with such outlying data.}
\label{fig:outlier}
\end{figure}

Matrix norm of the difference between networks is often used as a measure of similarity between networks  \citep{banks.1994,zhu.2014}. Given two networks $\mathcal{X}^1=(V, w^1)$ and $\mathcal{X}^2=(V, w^2)$, the $L_l$-norm of network difference is given by
$$
 D_l (\mathcal{X}^1,\mathcal{X}^2) = \parallel w^1 - w^2 \parallel_{l} =
            \Big(  \sum_{i,j}  \big| w^1_{ij} - w^2_{ij} \big|^{l}  \Big)^{1/l}. 
\label{eq:norm}
$$
Note $L_l$ is the element-wise Euclidean distance in $l$-dimension. When $l=\infty,$ $L_{\infty}$-distance is written as 
$$
 D_{\infty} (\mathcal{X}^1,\mathcal{X}^2) = \parallel w^1 - w^2 \parallel_{\infty} =
            \max_{\forall i,j}  \big| w^1_{ij} - w^2_{ij} \big|. 
\label{eq:inf_norm}
$$
The element-wise differences  may not capture additional higher order similarity. For instance, there might be relations between a pair of columns or rows  \citep{zhu.2014}. Also $L_1$ and $L_2$-distances usually surfer the problem of outliers. Few outlying extreme edge weights may severely affect the distance. Given two identical networks that only differ in one outlying edge with infinite weight (Figure \ref{fig:outlier}),  $L_{l}(\mathcal{X}^1, \mathcal{X}^2)= \infty$ and $L_{\infty}(\mathcal{X}^1, \mathcal{X}^2)= \infty$. Thus, the usual matrix norm based distance is sensitive to even a single outlying edge. However, topological distances are not sensitive to edges weights but sensitive to the underlying topology. Further, these distances ignore the underlying topological structures. Thus, there is a need to define distances that are more topological. 


\subsection{Log-Euclidean distance}

The log-Euclidean was previously used in measuring distance between correlation based brain networks \citep{qiu.2015,chung.2015.TMI}, which are supposed to reside in the space of positive definite symmetric (PDS) matrices. 
Let $\mathcal{S}_p$ be the space of  symmetric (PDS) matrices of size $p \times p$.
Let $\mathcal{P}_p$ be the space of positive definite symmetric (PDS) matrices. Note that the dimension of $\mathcal{S}_p$ and $\mathcal{P}_p$ is $p(p+1)/2$, i.e., $\mathcal{S}_p, \mathcal{P}_p \subset \mathbb{R}^{p(p+1)/2}$. While $\mathcal{S}_p$ is a flat Euclidean space, $\mathcal{P}_p$ is a curved manifold. Even though, various computation in  $\mathcal{P}_p$ is fairly involving, the use of exponential map $exp:\mathcal{S}_p \to   \mathcal{P}_p$ and its inverse make computations  in $\mathcal{P}_p$ tractable. The exponential map is realized with matrix exponential. 
In practice, the singular value decomposition is mainly used in computing the matrix exponential as follows. 
For $X \in \mathcal{S}_p$, there exists an orthogonal matrix $Q$ such that 
$$X=Q^\top \Lambda Q$$
with $\Lambda =diag(\lambda_1, \cdots, \lambda_p)$, the diagonal matrix consisting of entries $\lambda_1, \cdots, \lambda_p$.
Then matrix exponential is defined as
$$ exp X = Q^\top (exp \Lambda)Q.$$
Thus $exp X = Q^\top (exp \Lambda)Q$, where $exp \Lambda$ is the diagonal matrix consisting of entries $e^{\lambda_1}, \cdots, e^{\lambda_p}$. Similarly, for $Y \in \mathcal{P}_m$, we have decomposition $Y=Q^{\top} \Lambda Q $ with $\Lambda =diag(\lambda_1, \cdots, \lambda_p)$. Then the matrix logarithm of $Y$ is computed as 
$$log Y = Q^\top (\log \Lambda) Q.$$

If matrix $Y$ is nonnegative definite with zero eigenvalues, the matrix logarithm is {\em not} defined since $\log 0$ is not defined. Thus, we cannot apply logarithm directly to  rank-deficient large correlation and covariance matrices obtained from small number of subjects. One way of applying logarithm to nonnegative definite matrices is to make matrix $Y$ diagonally dominant by adding a diagonal matrix $\alpha I$ with suitable choice of relatively large $\alpha$ \citep{chan.1997}. Alternately, we can perform a graphical LASSO-type of sparse model and obtain the closest possible positive definite matrices \citep{chung.2015.TMI,qiu.2015,mazumder.2012}.

For $X, Y \in \mathcal{P}_p$, the log-Euclidean metric is given by
 the Frobenius norm \citep{qiu.2015}
$$
D_{LE}(X,Y) = \|\log X - \log Y \|_F\nonumber =  \sqrt {\mathrm{tr} \big( \log(X) - \log(Y) \big) ^2 }.
$$
The log-Euclidean distance was used in \cite{qiu.2015} to compute the mean of functional brain networks and the variation of each individual network around the mean. If $C_i$ denotes the brain network represented as the PDS correlation matrix of the $i$-th subject, the average of all brain network within log-Euclidean framework is given by
$$\bar{C} = \exp \Big(\frac{1}{n}\sum_{i=1}^n \log C_i \Big).$$
The metric  was also used  in the local linear embedding of brain functional networks \citep{qiu.2015} and regression for resting-state functional brain networks in the PDS space \citep{huang.2020.NM}.

\subsection{Graph matching}

Graph matching is well formulated established method for matching two different graphs via combinatorial optimization. The method is often used in distributed controls and computer vision  \citep{zavlanos.2008}. Suppose two graphs $\mathcal{X}_1 = (V, w_1)$ and $\mathcal{X}_2 = (V, w_2)$ with $p$ nodes are given. If $A_1$ and $A_2$ are adjacency matrices of $\mathcal{X}_1$ and $\mathcal{X}_2$, the graph matching cost function is given by
$$D_{GM} (\mathcal{X}_1, \mathcal{X}_2) = \min_{Q} \| Q A_1 Q ^{\top} -  A_2 \|_2^2,$$
where $Q$ is the permutation matrix that shuffles the node index. Two graphs are {\em isomorphic} if $D_{GM} =0$. All isomorphic graphs have the same structure, since one obtain $A_2$ from $A_1$ by relabelling of the nodes. 
However, not every two graphs are isomorphic. If $\lambda_1^i \geq \lambda_2^i \geq \cdots \geq \lambda_p^i$ are eigenvalues of $A_i$. Then we can show that  \citep{zavlanos.2008}
$$\| Q A_1 Q ^{\top} -  A_2 \|_2^2  \geq \sum_{i=1}^p (\lambda_i^1 - \lambda_i^2)^2,$$
which is the lower bound for the graph matching cost $D_{GM}$.

The main limitation of graph matching is the exponential run time $2^{\cal{O}(\sqrt{ p \log p})}$, which does not scale well for large $p$ \citep{babai.1983}. The other limitation is that if the size of node sets don't match, it is difficult to apply the graph matching algorithm. 
Additional nodes are argumented to match the node sets but the argumentation can be somewhat arbitrary \citep{guo.2020}.
The graph matching has been used in matching and averaging heterogenous tree structures such as brain artery trees and neuronal trees \citep{guo.2020}.  If the same brain parcellation is used across subjects, we do not need realignment of node labels so graph matching has not seen many applications in brain network analysis. However, with the availability of many different parcellations, it might be useful matching networks across different parcellations.

\subsection{Canonical correlations}

Many existing approaches for measuring distance between brain networks assume the size of networks to be the same. Otherwise, it is difficult to define distance and loss functions. The canonical correlation is perhaps one of few statistical similarly measure that enable to compute the distance between the measurements of different sizes  \citep{hotelling.1992}. It might be useful for comparing brain networks obtained parcellations of different sizes.

Given two vectors $X=(x_1, \cdots, x_n)^{\top}$ and $Y=(y_1, \cdots, y_m)^{\top}$, the canonical correlation between $X$ and $Y$ is given by 
$$\rho = \max_{a, b} corr( a^{\top} X, b^{\top} Y ).$$
Various numerical methods are available for computing $\rho$. It is computed using {\tt canocorr} in MATLAB and {\tt cancor} or {\tt CCA} in R package. For brain connectivity matrices, we vectorize only the upper triangle components of the connectivity matrices and compute the canonical correlations. Numerically, the canonical correlation is mainly computed vis the singular value decomposition (SVD). 

 The canonical correlation has been widely used in various applications including deep learning \citep{andrew.2013}, multiview clustering \citep{chaudhuri.2009}. In brain imaging, it is mainly used in correlating measurements from different modalities. In \citet{avants.2010}, fractional anisotropy values from diffusion tensor imaging and  cortical thickness from MRI were correlated using canonical correlations. In \citet{correa.2009}, fMRI, structural MRI and EEG measurements are correlated using canonical correlations.

\section{Preliminary: Persistent homology} 

We start with the basic mathematical understanding of persistent homology.

\subsection{Simplical homology}
\index{simplical homology}

A high dimensional object can be approximated by the point cloud data $X$ consisting of $p$ number of points. If we connect points of which distance satisfy a given criterion, the connected points start to recover the topology of the object. Hence, we can represent the underlying topology as a collection of the subsets of $X$ that consists of nodes which are connected \citep{edelsbrunner.2010, hart.1999}. Given a point cloud data set $X$ with a rule for connections, the topological space is a simplicial complex and its element is a simplex \citep{zomorodian.2009}. For point cloud data, the Delaunay triangulation is probably the most widely used method for connecting points. The Delaunay triangulation represents the collection of points in space as a graph whose face consists of triangles. Another way of connecting point cloud data is based on Rips complex often studied in persistent homology. 

Homology is an algebraic formalism  to associate a sequence of objects with a topological space \citep{edelsbrunner.2010}.  In persistent homology, the algebraic formalism is usually built on top of objects that are hierarchically nested such as morse filtration, graph filtration and dendrograms. Formally homology usually refers to homology groups which are often built on top of a simplical complex for point cloud and network data \citep{lee.2014.MICCAI}.

The $k$-simplex $\sigma$ is the convex hull of $v+1$ independent points $v_0, \cdots, v_k$.  A point is a $0$-simplex, an edge is a $1$-simplex, and a filled-in triangle is a $2$-simplex. A {\em simplicial complex} is a finite collection of simplices such as points (0-simplex), lines (1-simplex), triangles (2-simplex) and higher dimensional counter parts \citep{edelsbrunner.2010}. A {\em $k$-skeleton} is a simplex complex  of up to $k$ simplices. Hence a graph is a 1-skeleton consisting of $0$-simplices (nodes) and $1$-simplices (edges). There are various simplicial complexes. The most often used simplicial complex in persistent homology is the Rips complex.

\begin{figure}[t]
\centering
\includegraphics[width=1\linewidth]{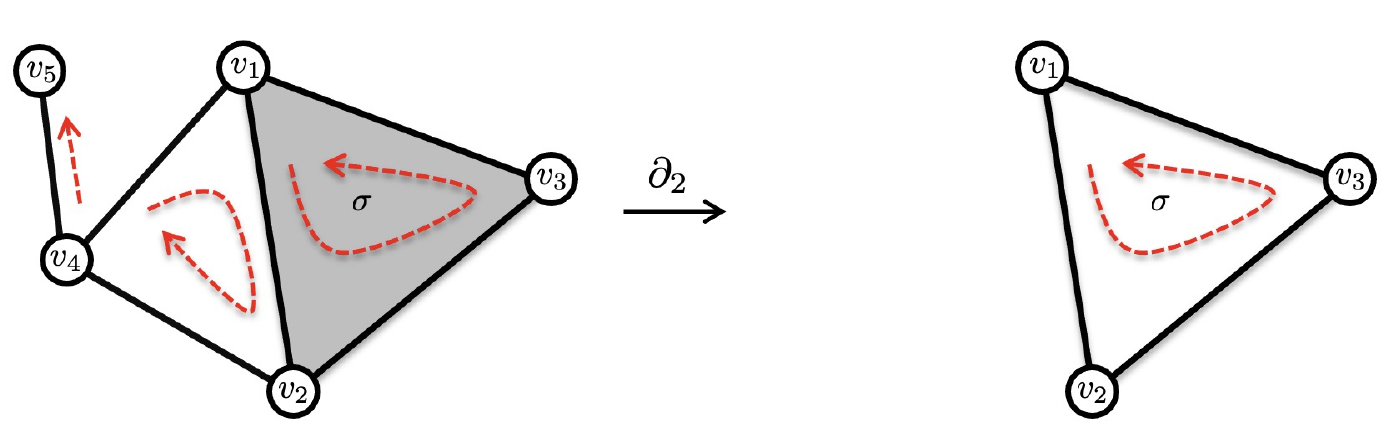}
\caption{A simplicial complex with 5 vertices and 2-simplex $\sigma =[v_1, v_2, v_3]$ with a filled-in face (colored gray). 
After boundary operation $\partial_2$, we are only left with 1-simplices $[v_1, v_2] + [v_2, v_3] + [v_3, v_1]$, which is the boundary of the filled in triangle. The complex has a single connected component ($\beta_0=1$) and a single 1-cycle. The dotted red arrows are the orientation of simplices.}
\label{fig:simplex}
\end{figure}

The  boundary operations have been very useful for effectively quantifying persistent homology. Let $C_k$ be the collection of $k$-simplices. We define the $k$-th boundary operator 
$$\partial_k: C_k \to C_{k-1}.$$
that removes the filled-in interior of $k$-simplices. Consider a filled-in triangle $\sigma = [v_1, v_2, v_3] \in C_2$ with three vertices $v_1, v_2, v_3$ in Figure \ref{fig:simplex}. 
The boundary operator $\partial_k$ applied to $\sigma$ resulted in the collection of three edges that forms the boundary of $\sigma$:
\bqn \partial_2 \sigma = [v_1,v_2] + [v_2,v_3] + [v_3,v_1] \in C_1. \label{eq: persist-p2} \eqn
If we give the direction or orientation to edges such that 
$$[v_3, v_1] = -[v_1, v_3],$$
and use edge notation $e_{ij} = [v_i, v_j]$, we can write (\ref{eq: persist-p2}) as
\bq \partial_2 \sigma = e_{12} + e_{23} + e_{31} = e_{12} + e_{23} - e_{13}. \eq

We can apply the boundary operation $\partial_1$ further to $\partial_2 \sigma$ and obtain 
\bq \partial_1 \partial_2 \sigma &=& 
\partial_1 e_{12}  + \partial_1 e_{23} + \partial_1 e_{31} \\
&=& v_2 - v_1 + v_3 - v_2 +  v_1 -v_3 =0.
\eq
The boundary operation twice will results in an empty set. Such algebraic representation for boundary operation has been very useful for effectively quantifying persistent homology.

The boundary operator $\partial_k$ can be represented using the \textit{boundary matrix}, which is  the higher dimensional version of the incidence matrix. \citep{lee.2018.ISBI,lee.2019.CTIC,schaub.2018} showing how $(k-1)$-dimensional simplices are forming $k$-dimensional simplex. The boundary matrix $\boldsymbol{\partial}_k$ is a matrix of size $n \times m$ consisting of ones and zeros, where $n$ is the total number of $(k-1)$-dimensional simplices and $m$ is the total number of $k$-dimensional simplices in the simplicial complex. Index the $(k-1)$-dimensional simplices as $\sigma_1, \cdots, \sigma_n$ and $k$-dimensional simplices as $\tau_1, \cdots, \tau_m$. The $(i,j)$-th entry of $\boldsymbol{\partial}_k$  is one if $\sigma_i$ is a face of $\tau_j$ otherwise zero. The entry can be -1 depending on the orientation of $\sigma_i$. 

 For the simplicial complex in Figure \ref{fig:simplex}, the boundary matrices are given by
\bq
\boldsymbol{\partial}_2 &=& 
\begin{array}{c}
\\
e_{12}\\
e_{23}\\
e_{31}\\
e_{24}\\
e_{41}\\
e_{45}
\end{array}
\begin{array}{c}
\sigma\\
\left(
\begin{array}{c}
 1 \\
 1   \\
 1 \\
 0\\
 0\\
 0 
\end{array}
\right)
\end{array}\\
\boldsymbol{\partial}_1 &=& 
\begin{array}{c}
\\
v_1\\
v_2\\
v_3\\
v_4\\
v_5
\end{array}
\begin{array}{c}
\begin{array}{cccccc}
e_{12} & e_{23} & e_{31} & e_{24} & e_{41} & e_{45}\\
\end{array}\\
\left(
\begin{array}{cccccc}
 -1 & 0 &  1  &  0&  1 & 0\\
 1 &  -1 & 0  &  -1&  0 &0\\
 0 &  1 & -1  &  0&  0 &0\\
 0 & 0& 0&  1&   -1 &-1\\
 0& 0& 0 & 0&  0 & 1
\end{array}
\right)
\end{array}\\
 \boldsymbol{\partial}_0 &=& 
\begin{array}{c}
\\
0
\end{array}
\begin{array}{c}
v_1 \;\; v_2 \;\; v_3 \;\;  v_4 \;\;  v_5\\
\left(
\begin{array}{ccccc}
 0 & 0 &  0 & 0 & 0
\end{array}
\right)
\end{array}.\\
\eq
Although we put direction in the boundary matrix $\boldsymbol{\partial}_1$ by adding sign, the Betti number $\beta_1$ computation will be invariant. With  boundary operations,  we can build a vector space $C_k$ using the set of $k$-simplices as a basis. 
 The vector spaces $C_k,$ $C_{k-1}, C_{k-2}, \cdots$ are then sequentially nested by boundary operator $\partial_k$ \citep{edelsbrunner.2010}:
 \bqn \cdots \xrightarrow{\partial_{k+1}} C_k \xrightarrow{\partial_k} C_{k-1} \xrightarrow{\partial_{k-1}} C_{k-2}  \xrightarrow{\partial_{k-2}}  \cdots. \label{eq:C_k} \eqn
 Such nested structure is called the {\em chain complex}. Let $B_k$ be a collection of boundaries obtained as the image of $\partial_k$, i.e., 
 $$B_k = \{ \partial_k \sigma : \sigma \in C_k \}.$$ Let $Z_k$ be a collection of cycles obtained as the kernel of $\partial_k$, i.e., 
 $$Z_k = \{ \sigma \in C_k: \partial_k \sigma =0  \}.$$ 
For instance, the 1-cycle formed by edges $e_{12}, e_{23}, e_{31}$ in Figure \ref{fig:simplex} is the boundary of the filled-in gray colored triangle $\sigma$. The boundaries $B_k$ form subgroups of the cycles $Z_k$, i.e, $B_k \subset Z_k$. We can partition  $Z_k$ into cycles that differ from each other by boundaries through the quotient space $$H_k= Z_k/B_k,$$ which is called the $k$-th homology group. The elements of the $k$-th homology group are often referred to as $k$-dimensional cycles or $k$-cycles. The $k$-th Betti number $\beta_k$ is then the number of $k$-dimensional cycles, which is given by the rank of $H_k$, i.e.,
\bqn \beta_k = rank (H_k) = rank (Z_k) - rank (B_k). \label{eq:betti} \eqn 
The 0-th Betti number is the number of connected components while 1-st Betti number is the number of cycles. 

The Betti numbers $\beta_k$ are usually algebraically computed by reducing the boundary matrix $\boldsymbol{\partial}_k$ to the Smith normal form, which has a block diagonal matrix as a submatrix in the upper left, via Gaussian elimination \citep{edelsbrunner.2010}. For instance, the boundary matrices $\boldsymbol{\partial}_i$ in Figure \ref{fig:simplex} is transformed to the Smith normal form $\mathcal{S}(\boldsymbol{\partial}_i)$  after Gaussian elimination as
$$
\mathcal{S}(\boldsymbol{\partial}_1)= \left(
\begin{array}{cccccc}
 1 & 0 &  0  &  0&  0 & 0\\
 0 &  1 & 0  &  0&  0 &0\\
 0 &  0 & 1  &  0&  0 &0\\
 0 & 0& 0&  1&   0 & 0\\
 0& 0& 0 & 0&  0 & 0
\end{array}
\right), \quad  \mathcal{S}(\boldsymbol{\partial}_2)= \left(
\begin{array}{c}
 1 \\
 0   \\
 0 \\
 0\\
 0\\
 0 
\end{array}
\right).
$$
In the Smith normal form  $\mathcal{S}(\boldsymbol{\partial}_k)$, the number of columns containing only zeros is $rank(Z_k)$, the number of $k$-cycles while the number of rows containing one is $rank(B_{k-1})$, the number of $(k-1)$-cycles that are boundaries.
From (\ref{eq:betti}), the Betti number computation involves the rank computation of two boundary matrices. In Figure \ref{fig:simplex} example, there are $rank(Z_1)=2$ zero columns and $rank (B_0) = 4$ non-zero rows. $rank(Z_0)=5$ is trivially the number of nodes in the simplicial complex while there are $rank(B_1) =1$ for $\mathcal{S}(\boldsymbol{\partial}_2)$. Thus, we have
\bq \beta_0 &=& rank (Z_0) - rank (B_0) =  5 - 4 = 1,\\
\beta_1 &=&  rank (Z_1) - rank (B_1) =   2 - 1 = 1.
\eq

The Betti numbers can be also computed using the Hodge Laplacian without Gaussian elimination. The standard graph Laplacian is defined as
$$\Delta_0 = \boldsymbol{\partial}_1  \boldsymbol{\partial}_1^{\top},$$
which is also called the 0-th Hodge Laplacian \citep{lee.2018.ISBI}. In general, the $k$-th Hodge Laplacian is defined as
$$\Delta_k =  \boldsymbol{\partial}_{k+1}  \boldsymbol{\partial}_{k+1}^{\top} + \boldsymbol{\partial}_k  \boldsymbol{\partial}_k^{\top}.$$
The boundary operation $\partial_{k}$ only depends on $k$-simplices. Thus, $\Delta_k$ is uniquely determined by $(k+1)-$ and $k$-simplices. The $k$-th Laplacian is sparse a $n_k \times n_k$ positive semi-definite symmetric matrix, where $n_k$ is the number of $k$-simplices  in the network \citep{friedman.1998}. Then the $k$-th Betti number $\beta_k$ is the dimension of $ker \Delta_k$, which is given by computing the rank of $\Delta_k$. The 0th Betti number, the number of connected component, is computed from $\Delta_0$ while the 1st Betti number, the number of cycles, is computed from $\Delta_1$. 

The $k$-th hodge Laplacian depends only on $n_k$  number of $k$-simplices in the data. After lengthy algebraic derivation, we can show that
$$\Delta_k = D - A_u + (k+1)I_{n_k} + A_l,$$
where  $A_u$ and $A_l$ are the upper and lower adjacency matrices between the $k$-simplices. 
$D=diag(deg(\sigma_1), \cdots, deg(\sigma_{n_k}))$ is the diagonal matrix consisting of the sum of node degrees of simplices $\sigma_j$ \citep{muhammad.2006}.

\subsection{Rips filtrations: filtrations on point cloud data} 
\index{Rips complex!triangulation}
\index{Rips}
\index{filtrations!Rips}
\index{filtration}

\begin{figure}[t]
\begin{center}
\includegraphics[width=1\linewidth]{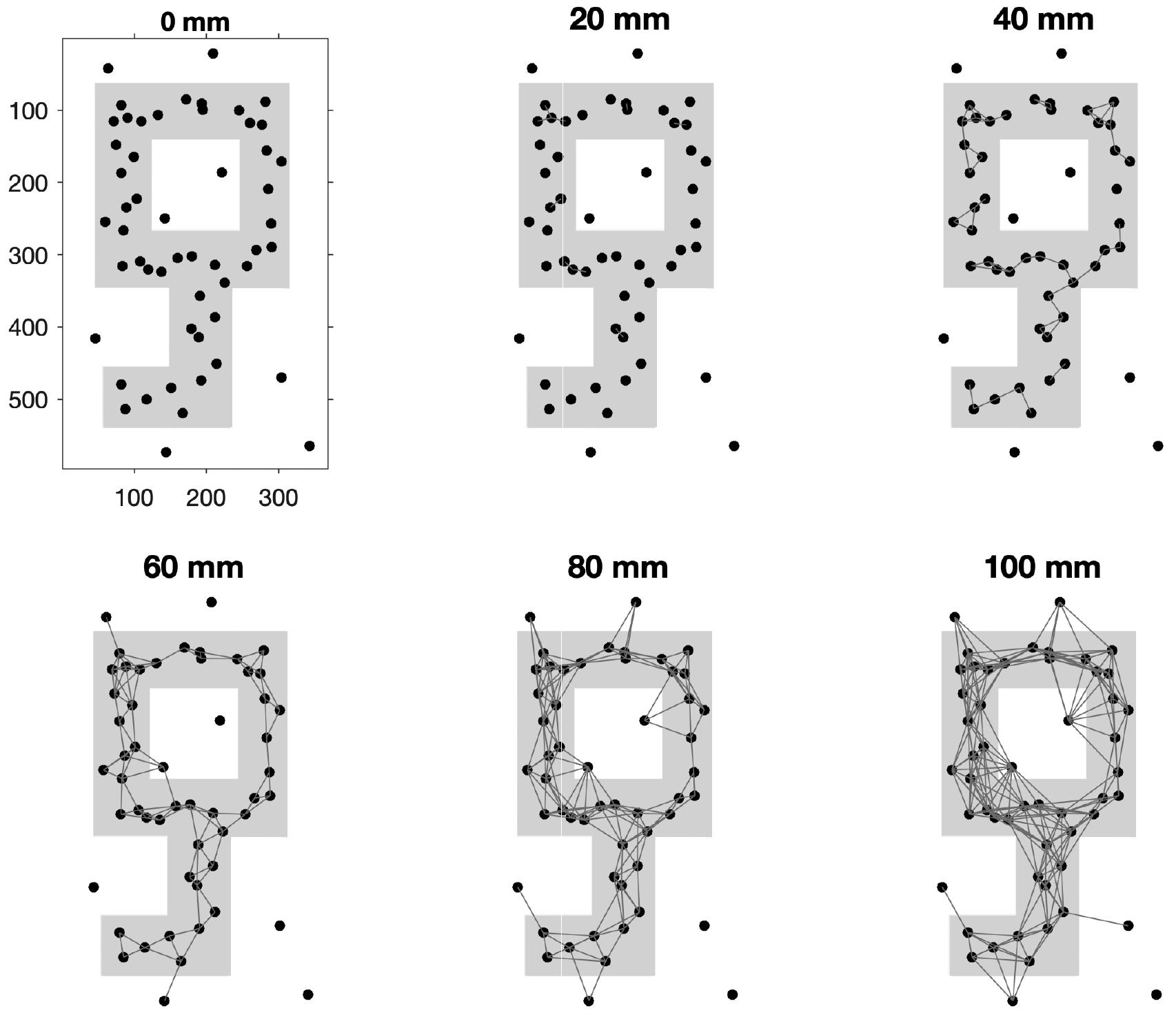}
\caption{Rips filtration on 1-skeleton (Rips complex consisting of only nodes and edges) of the point cloud data that was sampled along the underlying key shaped data. If two points are within the given radius, we connect them with an edge.}
\label{fig:topology-ripskey}
\end{center}
\end{figure}

The Rips complex has been the main building block for persistent homology  and defined on top of the point cloud data \citep{ghrist.2008}. The {\em Rips complex} is a  simplicial complex constructed by connecting two data points if they are within specific distance $\epsilon$. Figure \ref{fig:topology-ripskey} shows an example of the Rips complex that approximates the gray object with a point cloud. Given a point cloud data, the Rips complex $R_{\epsilon}$ is a simplicial complex whose $k$-simplices correspond to unordered $(k+1)$-tuples of points which are pairwise within distance $\epsilon$ \citep{ghrist.2008}. While a graph has at most 1-simplices, the Rips complex has at most $k$-simplices. The Rips complex has the property that
$$\mathcal{R}_{\epsilon_0} \subset \mathcal{R}_{\epsilon_1}\subset \mathcal{R}_{\epsilon_2}\subset \cdots $$
for $0=\epsilon_{0} \le \epsilon_{1} \le \epsilon_{2} \le \cdots.$
When $\epsilon=0$, the Rips complex is simply the node set $V$. By increasing the filtration value $\epsilon$, we are connecting more nodes so the size of the edge set increases.
Such the nested sequence of the Rips complexes is called a {\em Rips filtration}, the main object of interest in the persistent homology \citep{edelsbrunner.2008}. The increasing $\epsilon$ values are called the filtration values. 

One major problem of the Rips complex is that as the number of vertices $p$ increase, the resulting simplical complex becomes very dense. Further, as the filtration values increases, there exists an edge between ever pair of vertices and filled triangle between every triple of vertices. At higher filtration values, Rips filtration becomes very ineffective representation of data.

\subsection{Morse filtrations and elder rule}

\begin{figure}[t]
\begin{center}
\includegraphics[width=1\linewidth]{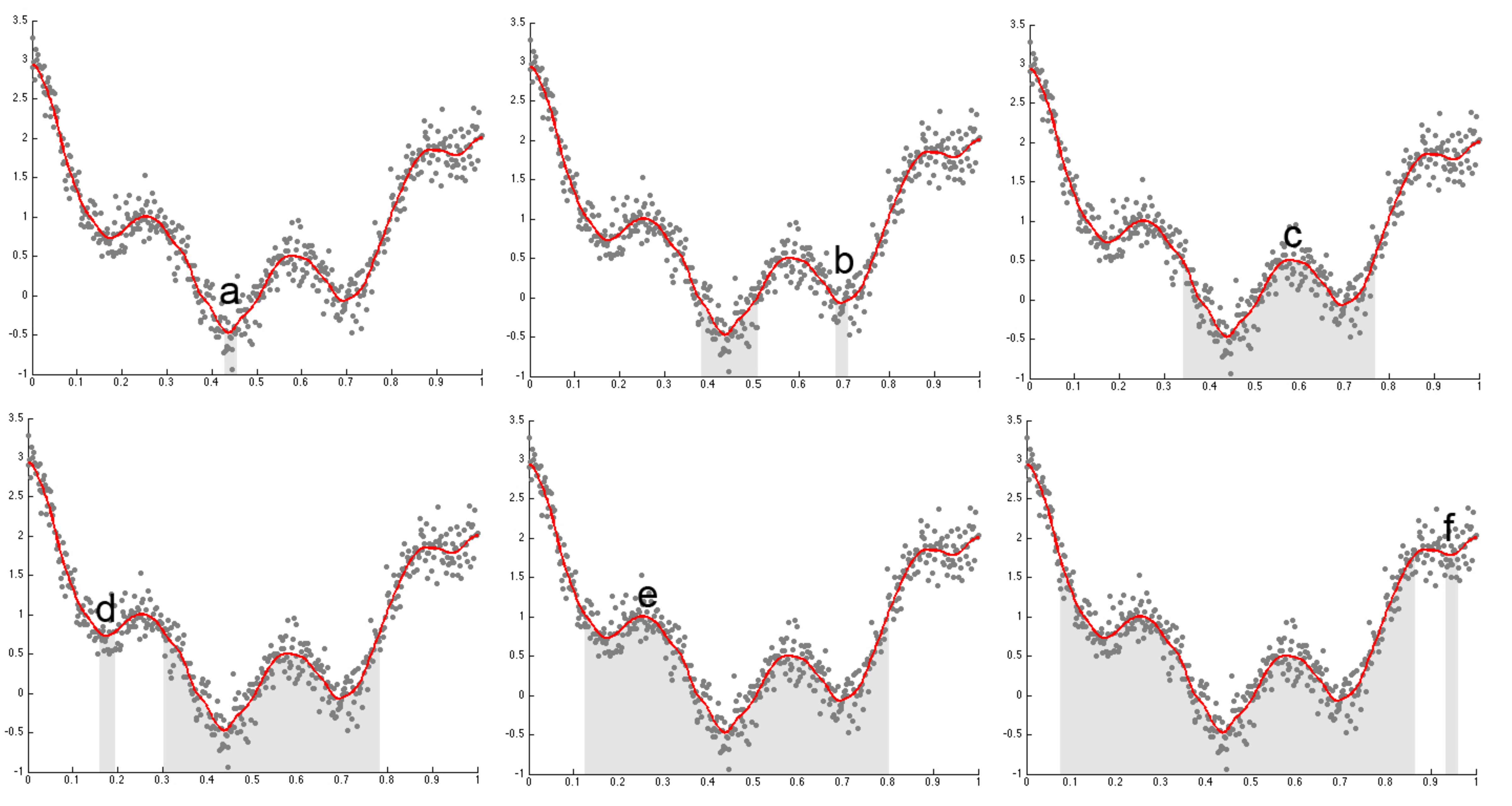}
\caption{\label{fig:bd} The births and deaths of connected components  in the sublevel sets in a Morse filtration \citep{chung.2009.IPMI}. We have local minimums $a < b < d <f$ and local maximums $c< e$. At $y=a$, we have a single connected component (gray area). As we increase the filtration value to $y=b$, we have the birth of a new component (second gray area). At the local maximum $y=c$, the two sublevel sets merge together to form a single component. This is viewed as the death of a component. The process continues till we exhaust all the critical values. Following the Elder rule, we pair birth to death: $(c,b)$ and $(e,d)$.  Other critical values are paired similarly. These paired points form the persistent diagram (PD).}
\end{center}
\end{figure}

A function is called a {\em Morse function} if all critical values are distinct and non-degenerate, i.e., the Hessian does not vanish \citep{milnor.1973}. For a 1D Morse function $y=f(t)$, define sublevel set $R(y)$ as 
$$R(y) = f ^{-1}(-\infty, y] = \{t\in\mathbb{R}: f(t)< y\}.$$ The sublevel set is the domain of $f$  satisfying $f(t) \leq y$. As we increase height $y_1 \leq y_2 \leq  y_3 \leq\cdots$, the sublevel set gets bigger such that
$$R(y_1) \subset R(y_2) \subset R(y_3) \subset \cdots.$$
The sequence of the sublevel sets form a {\em Morse filtration} with filtration values $y_1, y_2, y_3, \cdots$. 
Let $\beta_0(y)$ be the $0$-th Betti number of $R(y)$. $\beta_0(y)$ counts the number of connected components in 
$R(y)$. The number of connected components is the most often used topological invariant in applications \citep{edelsbrunner.2008}. $\beta_0(y)$ only changes its value as it passes through critical values (Figure \ref{fig:bd}). The birth and death of connected components in the Morse filtration is characterized by the pairing of local minimums and maximums. For 1D Morse functions, we do not have higher dimensional topological features beyond the connected components. 

Let us denote the local minimums as $g_1, \cdots, g_m$ and the local maximums as $h_1, \cdots, h_n$. Since the critical values of a Morse function are all distinct, we can combine all minimums and maximums and reorder them from the smallest to the largest: 
We further order all critical values together and let
$$g_1 = z_{(1)} < z_{(2)} < \cdots < z_{(m+n)} = h_n,$$
where $z_{i}$ is either $h_i$ or $g_i$ and $z_{(i)}$ denotes the $i$-th largest number in $z_1, \cdots, z_{m+n}$.  In a Morse function, $g_1$ is smaller than $h_1$ and $g_m$ is smaller than $h_n$ in the unbounded domain $\mathbb{R}$ \citep{chung.2009.IPMI}.

By keeping track of the birth and death of components, it is possible to compute topological invariants of sublevel sets such as the $0$-th Betti number $\beta_0$ \citep{edelsbrunner.2008}. As we move $y$ from $-\infty$ to $\infty$, at a local minimum, the sublevel set adds a new component so that 
$$\beta_0(g_i - \epsilon) = \beta_0( g_i ) +1$$
for sufficiently small $\epsilon$. This process is called the {\em birth} of the component. The newly born component is identified with the local minimum $g_i$. 

Similarly for at a local maximum, two components are merged as one so that
$$\beta_0(h_i - \epsilon) = \beta_0( h_i ) -1.$$
This process is called the {\em death} of the component. Since the number of connected components will only change if we pass through critical points and we can iteratively compute $\beta_0$ at each critical value as
$$\beta_0(z_{(i+1)}) = \beta_0(z_{(i)}) \pm 1.$$
The sign depends on if $z_{(i)}$ is maximum ($-1$) or minimum ($+1$). This is the basis of the Morse theory \citep{milnor.1973} that states that the topological characteristics of the sublevel set of Morse function are completely characterized by critical values.

To reduce the effect of low signal-to-noise ratio and to obtain smooth Morse function, either spatial or temporal smoothing have been often applied to brain imaging data before persistent homology is applied. In \citet{chung.2015.TMI, lee.2017.HBM}, Gaussian kernel smoothing was applied to 3D volumetric images. In \citet{wang.2018.annals}, diffusion was applied to temporally smooth data.

 \begin{figure}[t]
\includegraphics[width= .9\linewidth]{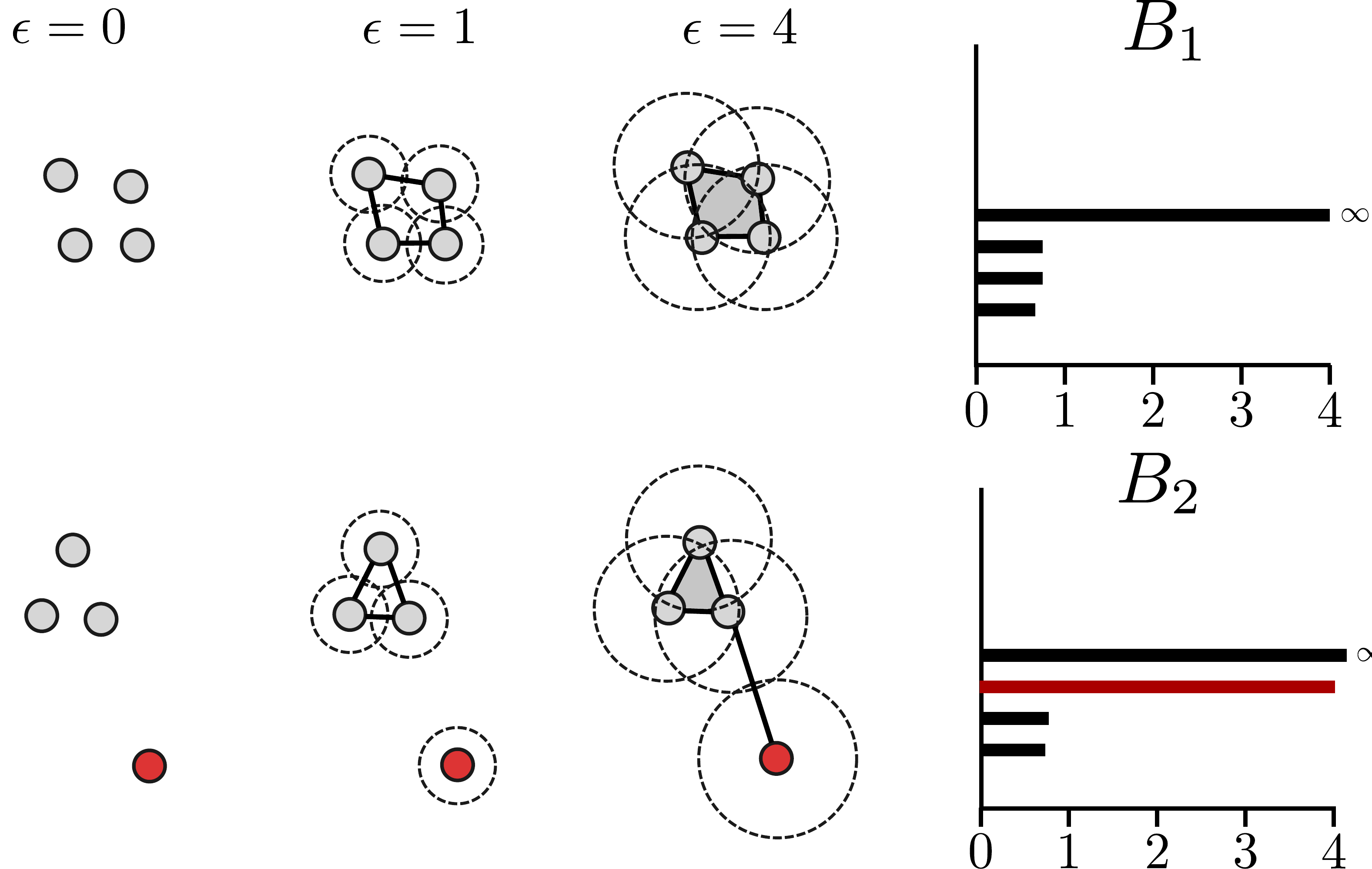}
\caption{A comparison of the barcodes obtained from a sample with no measurement error (top) and
a sample containing a single measurement error  (red point). While both point clouds may be taken from the same population, the presence of a single measurement error can dramatically impact the resulting barcodes. 
$B_1$ has most features with approximately equal lifetimes ($\tau_i - \xi_i$), where as $B_2$ has a feature with a large lifetime corresponding to the improperly measured point. Simply looking at the barcodes, it is difficult to differentiate topological signal to noise.}
\label{fig:exp}
\end{figure}

\subsection{Persistent diagrams \& barcodes}
\index{persistent diagrams}

 In persistent homology, the topology of underlying data can be represented by the birth and death of cycles. In a graph, the 0D and 1D cycles are a connected component and a cycle \citep{carlsson.2008.JCV}. During a filtration, cycles in a homology group appear and disappear.  If a  cycle appears at birth value $\xi$ and disappears at death value $\tau,$ it can be encoded into a point, $(\xi,\tau)  \; (0 \le \xi \le \tau < \infty)$ in $\mathbb{R}^2$. If $m$ number of cycles appear during the filtration of a network $\mathcal{X}=(V,w)$, the homology group can be represented by a point set $$\mathcal{P} (\mathcal{X}) = \left\{ (\xi_{1},\tau_{1}), \dots, (\xi_{m},\tau_{m}) \right\}.$$ 
This scatter plot is called the persistence diagram (PD) \citep{cohensteiner.2007}. A {\em barcode} encodes the birth time $\xi_i$ and death time $\tau_i$ of the cycle as an interval $(\xi_i,\tau_i)$. The PD and barcode encode topologically equivalent information. The length of a bar $\tau_i - \xi_i$ is the {\em persistence} of the cycle (Figure \ref{fig:exp}) \citep{edelsbrunner.2010}.  Longer bars corresponds to more persistent cycles that are considered as topological signal while shorter bars correspond to topological noise. However, recently such interpretation has been disputed and it may possible to have meaningful signal even in shorter bars \citep{bubenik.2020}.

\begin{figure}[t]
\centering
\includegraphics[width=1\linewidth]{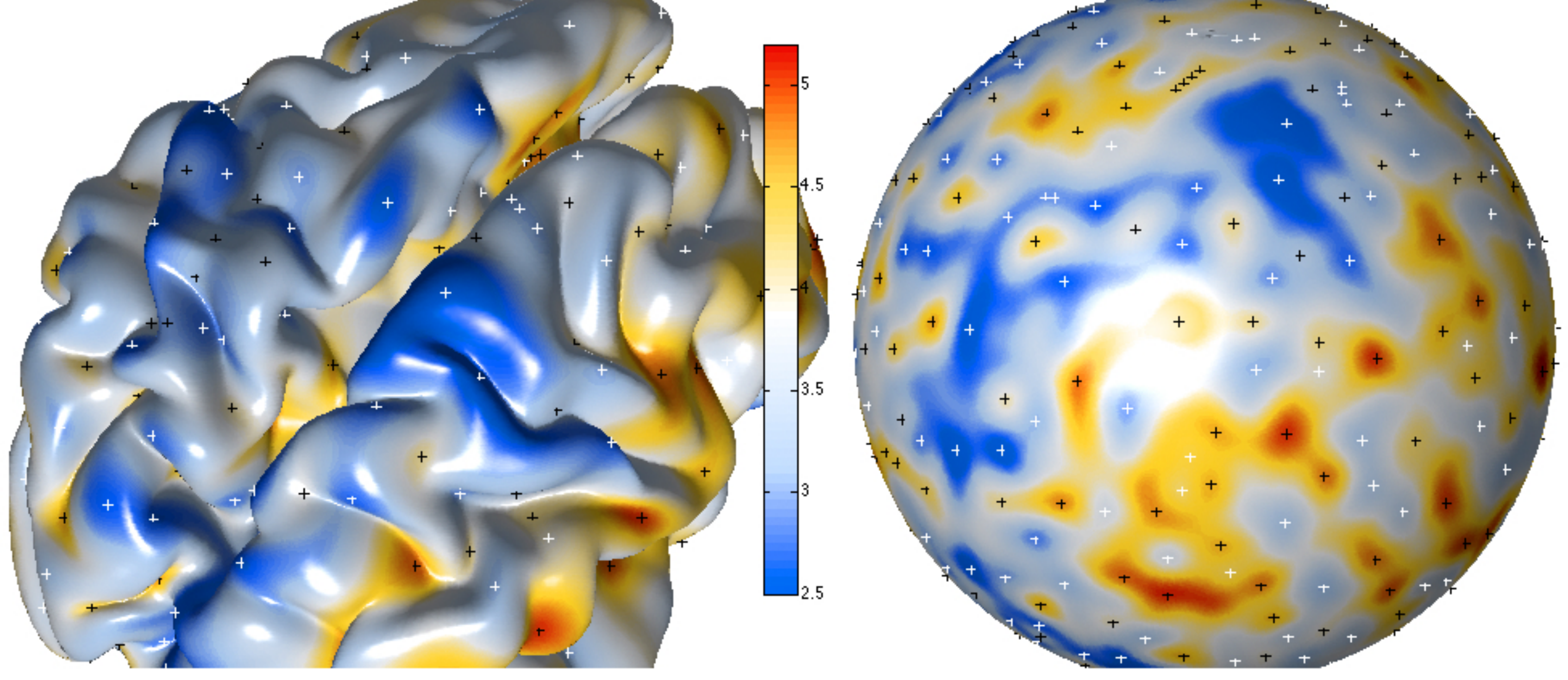}
\caption{Heat kernel smoothing was performed on cortical thickness to obtain smooth Morse function and projected onto a sphere 
\citep{chung.2009.MICCAI}. The critical points are identified by comparing coritcal thickness around each mesh vertex. 
The white (black) crosses are local minimums (maximums). They will be paired following Elders rule to obtain the reduced PD.}
\label{fig:minmax}
\end{figure}

The birth and death of connected components in the sublevel set of a function can be quantified and visualized by persistent diagram (PD) (Figure~\ref{fig:bd}). Consider a smooth Morse function $y=f(x)$ with unique critical values $a < b < c < d < e < f$ that estimate the underlying data. Such function can be estimated using kernel smoothing methods \citep{wang.2018.annals}. Then consider the Morse filtration that swaps $y$ value from $-\infty$ to $\infty$. Each time the horizontal line touches the local minimum $a < b < d < f$, a new component that contain the local minimum is born. Each time the line touches the local maximum $c < e$, the two components in the sublevel set merge together. This is considered as the death of a component. Following the {\em Elder Rule}, when we pass a maximum and merge two components, we pair the maximum (birth) with the higher of the minimums of the two components (birth) \citep{edelsbrunner.2008,edelsbrunner.2010,zomorodian.2005}. Doing so we are pairing the birth of a component to its death. Such paired points $(c,b)$ and $(e,d)$ form a {\em persistence diagram}. In general, PD can be algebraically represented as the weighted sum of Dirac delta functions
\bqn I(x,y) =  \sum_{i=1}^m c_i \delta(x-\xi_i, y - \tau_i) \label{eq:delta} \eqn
 with $\sum_{i=1}^m c_i =1$ such that $I(x,y)$ becomes a 2D probability distribution satisfying 
 $$  \int_{0 < x} \int_{y > x}    I(x,y) \; dx dy =1.$$
 Such probabilistic representation enables us to the access the vast library of distance and similarity measures between probability distributions. 
 Given 1D functional signal $f$ observed at time points $t_1,$ $\cdots$ $t_n$, the critical points can be numerically estimated obtained by checking sign changes in the finite differences $f(t_{i+1})-f(t_i).$
The estimation of critical points in higher dimensional functions are done similarly by checking the signs of neighboring voxels or nodes (Figure \ref{fig:minmax}) \citep{osher.2003,chung.2009.IPMI}.  

For higher dimensional Morse functions, saddle points can also create or merge sublevel sets so we also have to be concerned with them. Since there is no saddle points in 1D Morse functions, we do not need to worry about saddle points in 1D functional signals. The addition of the saddle points makes the construction of the persistence diagrams much more complex.  However, the saddle points do not yield a clear statistical interpretation compared to local minimums and maximums. In fact, there is no statistical methods or applications developed for saddle points in literature. Thus, they are often removed in the {\em reduced Morse filtration} \citep{chung.2009.MICCAI,pachauri.2011}. 

The use of critical points and values within  image analysis and computer vision has been relatively limited so far, and typically appear as part of simple image processing tasks such as feature extraction and edge detection  \citep{antoine.1996,cootes.1993,sato.1998}. The first or second order image derivatives may be used to identify the edges of objects to serve as the contour of an anatomical shape. In this context, image derivatives are computed after image smoothing and thresholded to obtain edges and ridges of images, that are used to identify voxels likely to lie on boundaries of anatomical objects. Then, the collection of critical points are used as a geometric feature that characterize anatomical shape. Specific properties of critical values as a topic on its own, however, has received less attention so far. One reason is that it is difficult to construct a streamlined linear analysis framework using critical points or values. In brain imaging, on the other hand, the use of extreme values has been quite popular  in the context of multiple comparison correction using the random field theory \citep{worsley.1996, taylor.2008,kiebel.1996}. In the random field theory, the extreme value of a statistic is obtained from, and is used to compute the $p$-value for correcting for correlated noise across neighboring voxels.

\section{Why statistical analysis in TDA hard?}
 The persistent diagrams and equivalent barcodes are the original most often used descriptors in persistent homology. However,  due to the heterogenous nature of PD, statistical analysis of PD and barcodes have been very difficult. 
 It is not even clear how to average PDs, the first critical step in building statistical frameworks. Even the Fr\'{e}chet mean is not unique, rendering it a challenging statistical issue to perform inference on directly on PDs \citep{bubenik.2015,chung.2009.IPMI,heo.2012}. To remedy the problem various methods including persistent landscape  \citep{bubenik.2015} and persistence images are proposed.

\subsection{Accumulating barcodes}

Even though barcodes possess similar stability properties as PD \citep{bauer2013induced}, the statistical analysis of barcodes is difficult. The difficulty is due to the heterogeneous algebraic representation as an unordered multiset of intervals 
$$B=\{ (\xi_1,\tau_1), (\xi_2,\tau_2), \cdots, (\xi_m,\tau_m) \},$$ 
where  $\xi_i$ and $\tau_i$ represent the birth and death times of the $i$-th topological features. Barcodes may not match across different networks. Also performing averaging, and subsequently constructing test statistics is more difficult as there is no unique Fr\`{e}chet mean in the space of barcodes \citep{kalivsnik2019tropical,adcock2013ring,hofer2019learning,turner2014frechet}.

Another important aspect is the inability of the barcode representation, and persistent homology in general, to distinguish between topological features that are intrinsic to the data or noises \citep{blumberg2014robust}.  Consider  a point cloud data coming from some measurements (Figure \ref{fig:exp}). The four points on top are all clustered together. On the other hand, one measurement (red point) can be far from the cluster as the result of measurement error. The noisy point produces a connected component with 
large persistence making the barcodes ($B_2$) very different from the barcodes ($B_1$) from the data with no measurement error.

Due to all these limitations, barcodes are often made into summary statistics such as 
the minimum, maximum and average length of a collection of barcodes \citep{pun2018persistent}. The \textit{accumulated persistence function} (APF) 
 accumulates the sum of the barcodes into a scalar value:
$$
APF(t) = \sum_{i:\mu_i\leq t}(\tau_i-\xi_i),
$$
where $\mu_i = (\tau_i + \xi_i)/2$ is the center of the $i$-th bar. Since APF is monotonically increasing over parameter $t$, it may be possible to construct KS-distance between APF and perform the exact topological interference \citep{chung.2017.IPMI}. APF was used in the  analysis of brain artery trees \citep{biscio2019accumulated} and determining if rs-fMRI is topologically stationary over time \citep{ssong.2020.ISBI}.

Barcodes can also be quantified in terms of their entropy $E(B)$, which is another summary statistic \citep{atienza2018stability}. Let
$l_i = \tau_i - \xi_i$ be the length of $i$-th barcode and $L = \sum_{i}l_i$ be the total length of all the barcodes.
Define the fraction of length of the $i$-th barcode $p_i  = l_i/L$ over the total length. Then the entropy $E(B)$ is defined as
\begin{equation}
E(B) = -\sum_{i}p_i \log p_i 
\label{eq:ent}
\end{equation}
This provides a single scalar measurement of the topological disorder of a given system \citep{atienza2018stability}. The barcode entropy has also been shown to outperform graph theoretic entropies, such as connectivity entropy and  Von Neumann entropy when characterizing complex networks and has been used in identifying pre-ictal and ictal states in epilepsy patients \citep{rucco2016characterisation, merelli2015topological,merelli2016topological}.

\begin{figure}[t]
\includegraphics[width= .9\linewidth]{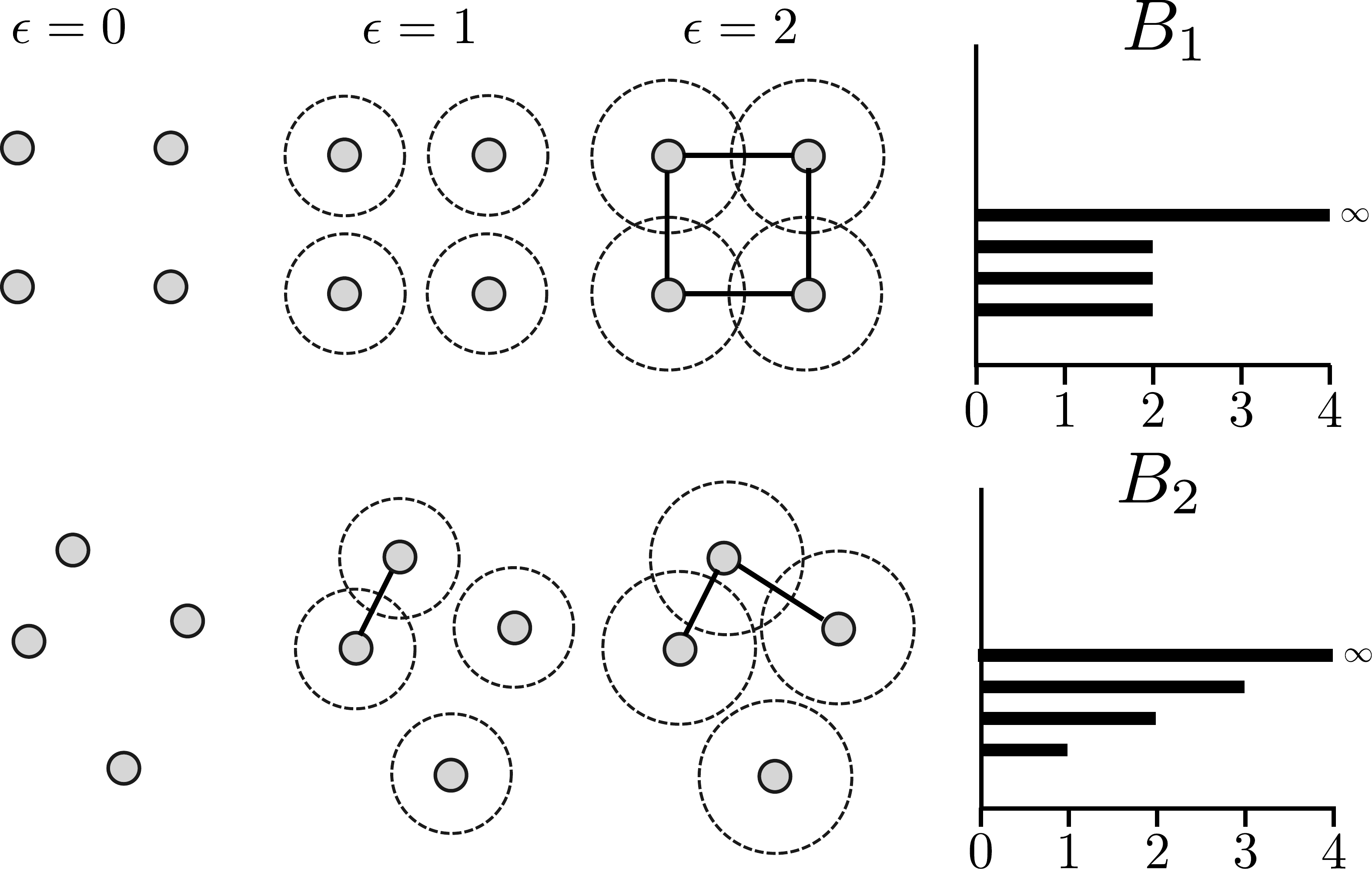}
\caption{A comparison of the barcodes for an ordered set of points (top) and a relatively unordered set of points (bottom). In the structured system, we find that most components have equal birth ($\xi_i$) and death ($\tau_i$) times, which result in equal lifetimes for each topological feature $(\tau_i - \xi_i)$, resulting in a high persistence entropy $E(B_1)$. For unordered system $B_2$, the components have different death points. Thus unordered system will have a lower persistence entropy, i.e., $E(B_1) > E(B_2)$.
}
\label{fig:ent}
\end{figure}

Figure \ref{fig:ent} illustrates how entropy differs for two systems. For ordered system $B_1$, ignoring the component that dies at $\infty$, we have $L = 6, p_1 = p_2 = p_3 = 2/6$ and  $E(B_1) = 1.1$. For unordered system$B_2$, we have $L = 6, p_1  = 1/6, p_2 = 2/6, p_3 = 3/6$ and$E(B_2) = 1.0$. Thus, ordered system has higher persistent entropy $E(B_1) > E(B_2)$. It would be investing to investigate if brain networks will exhibit lower persistence entropy.

\subsection{Persistent landscape}

The PD and equivalent barcodes are the original most often used descriptors in persistent homology. 
Even though, the PD  possess desirable bounding properties such as Lipschitz stability with respect to the bottleneck distance, statistical analysis on persistent PD is difficult since it is not a clear cut to define the average PD. Further the Fr\'{e}chet mean is not unique in PD, rendering it a challenging statistical issue to perform inference directly on PD \citep{chung.2009.IPMI, heo.2012}. To remedy the problem, persistent landscape was first proposed in \citet{bubenik.2015}, which maps PD to a Hilbert space.  

Given a barcode $(\xi_i,\tau_i)$, we can define the piecewise linear bump function
$h_i$ as
\bqn \label{eq: bump}
h_{i}(\epsilon)=\max ( \min (\epsilon - \xi_i, \tau_i -\epsilon),0). 
\eqn
The geometric representation of the bump function
\eqref{eq: bump} is a right-angled isosceles triangle with height equal to half of the base of the corresponding
interval in the barcode \citep{wang.2019.ICASSP,bubenik.2020.PL}. Given barcodes $(\xi_1, \tau_1), \cdots, (\xi_m, \tau_m)$, the persistent landscape $\lambda_k(\epsilon)$ is defined as
$$\lambda_k(\epsilon) = k\mbox{-th largest value of }  h_{i}(\epsilon).$$

With this new representation, it is possible to have a unique mean landscape $\bar \lambda(\epsilon)$ over subjects, which is simply the average of persistent landscape at each filtration value $\epsilon$. Such operation enables to construct the usual test statistic. The persistent landscapes have been applied to epilepsy EEG \citep{wang.2018.annals,wang.2019.ICASSP}, functional brain networks in fMRI \citep{stolz.2017,stolz.2018} and neuroanatomical shape modeling in MRI  \citep{garg.2017}. However, since scatter points in PD are converted to piecewise linear functions, statistical analysis on persistent landscape is not necessarily any easier than before. Thus, additional transformations on persistent landscapes is needed for statistical analysis. Since $\lambda_1(\epsilon) > \lambda_2(\epsilon) > \lambda_3(\epsilon) > \cdots$ at each fixed $\epsilon$, it is possible to build a monotone sequence of scalar values by computing area under function $\lambda_k(\epsilon)$ and perform the exact topological inference \citep{wang.2019.ICASSP}.

\subsection{Smoothing PD}

Due to the heterogenous nature of persistent diagrams (PD), it is difficult to even average PDs and perform statistical analysis. 
Starting with \citet{chung.2009.IPMI}, various smoothing methods were applied to PD such that  averaging and statistical analysis can be directly performed on the smoothed PD. \citet{chung.2009.IPMI}  discretized PD using  the the uniform square grid. A concentration map is then obtained by counting the number of points  in each pixel, which is equivalent to smoothing PD with a uniform kernel.  Notice that this approach is somewhat similar to the voxel-based morphometry \citep{ashburner.2000}, where brain tissue density maps are used as a shapeless metric for characterizing concentration of the amount of tissue. In \citet{reininghaus.2015},  persistence scale-space kernel approach was proposed, where points in PD is treated as heat sources modeled as Dirac-delta and diffusion is run on PD. Ignoring symmetry across $y=x$, diffusion of heat sources is given by the convolution of Gaussian kernel 
$$K_{\sigma}(x,y) = \frac{1}{2\pi \sigma^2} e^{-(x^2 + y^2)/ (2\sigma^2)}$$ on Dirac-delta functions as 
$$f(x,y) =\sum_{i=1}^m  K_{\sigma} * \delta(\xi_i, \tau_i) = \sum_{i=1}^m \frac{1}{2\pi \sigma^2} e^{- [(x- \xi_i)^2 + (y- \tau_i)^2 ]/ (2 \sigma^2)}.$$
This representation is known as the \textit{persistence surface}  \citep{adams.2017}.
\begin{figure}
  \includegraphics[width=\linewidth]{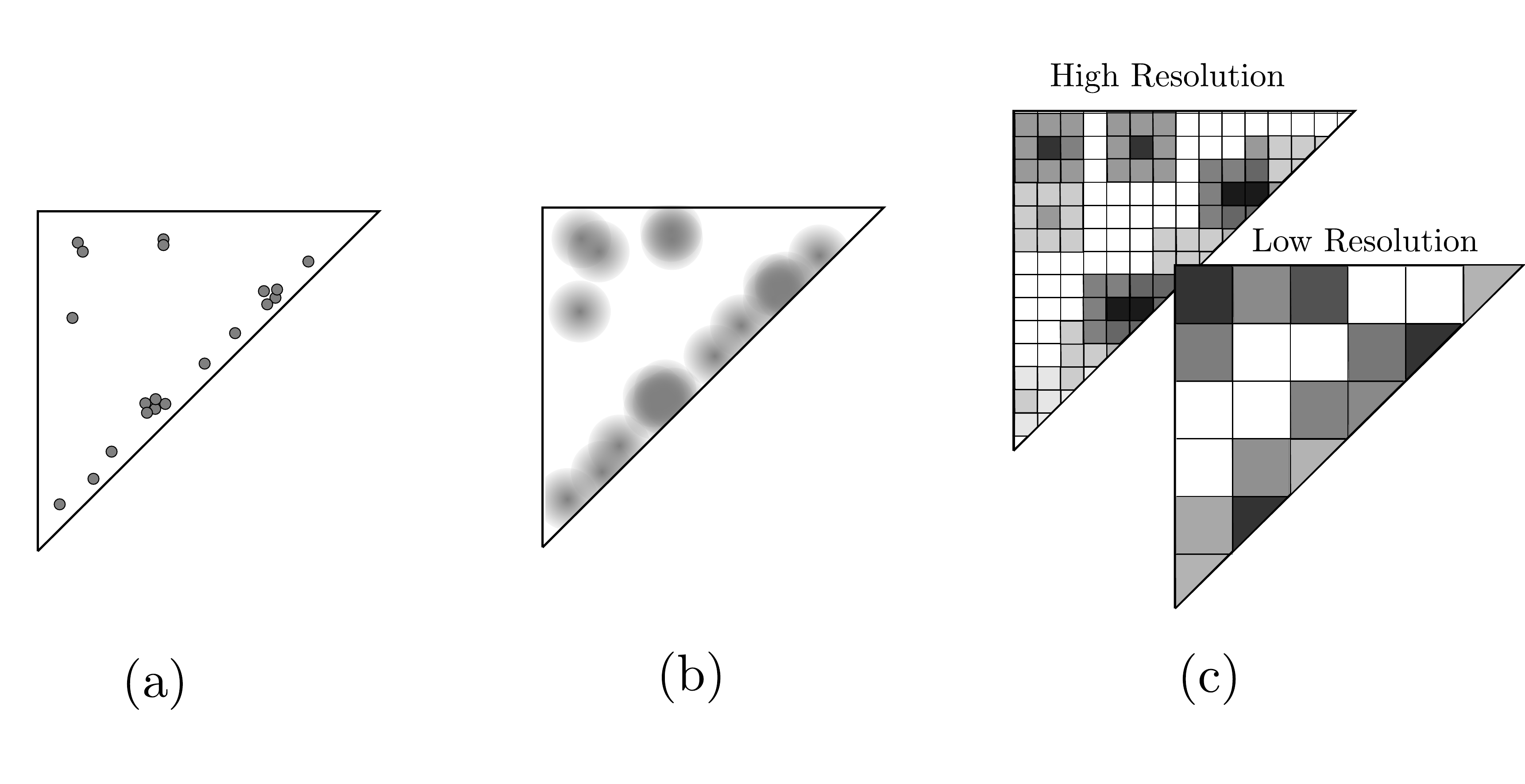}
  \caption{(a) Persistence Diagram (PD) scatter plot. (b) Persistence Surface. (c) Persistence Image (PI) at multiple grid resolutions. An illustration of the process for converting a PD into a PI representation. The scatter plot presented in (a) is fit with multiple 2-dimensional Gassians centered at each point in the PD, known as the persistence surface (b). The persistence surface is then discretized into a persistence image (c) by overlaying a grid, and assigning each grid square a scalar value equal to the integral of the sum of the Gaussians within the bounds of the grid square.}
  \label{fig:PI}
\end{figure}
Heat diffusion $f(x,y)$ is then discretized by integrating the function within each square grid of given resolution. An example of the discretization  is given in Figure \ref{fig:PI}-c. A \textit{persistence image} (PI) is simply the vector of square grid values.  There exists a stability result bounding the $L_2$-distance between PI by the bottleneck distance $D_B$ or Wasserstein distance $D_W$ on PD  \citep{adams.2017}:
\begin{equation}
   ||PI-PI'||_{2} \leq L\cdot D(PD,PD')
	\label{eq:imagestable}
\end{equation}
for some constant $L$. 
This is consequence of kernel smoothing, which reduces the spatial variability. The vectorized representation enables the PI to be used in various statistical and machine learning applications \cite{chazal2017introduction}. PI was used in the functional brain network study on Schizophrenia \citep{stolz.2018} and and in conjunction with deep learning for autism classification \citep{rathore2019autism}. The major theme across these applications is that the PI presents a simple, vectorized representation of the persistence diagram that can be readily used in many statistical and machine learning algorithms \citep{pun2018persistent}. 
The PI can be easily manipulated by adjusting the size of grid as well as the bandwidth $\sigma$ of kernel. 
This allows users to adjust the amount of information needed to represent PD in multiple scales for application needs.

One major drawback of the method is the inherent loss of information from both the grid discretization and the smoothing of PD. This loss of information can be detrimental to the analysis of the given dataset as important features may be over smoothed or unintentionally grouped together. One method of combatting this potential loss of information is to utilize weighting function $w(\xi,\tau)$ on the PI representation to emphasize certain features \citep{divol2019choice, adams.2017}. A popular method used is to assign linear or exponential weight such as $w(\xi,\tau) = (\tau-\xi)$ for each point in the PD that increases in correlation with the distance from the diagonal of the persistence diagram  before diffusion via the Gaussian kernel. This weights more toward the features of higher persistence making them robust to the subsequent analysis.

\section{Big data, scalability and approximation}

\subsection{Computational complexity}
The huge volumes of data in various imaging fields including neuroimaging and cosmology necessitate efficient algorithms for performing scalable TDA computation. The computing needs in these diverse fields share a lot of similarities and we can adapt many scalable methods developed in other areas to brain imaging data. In this section, we review some of the scalable schemes for simplifying big data computations. There are several ways in which TDA can help with big data challenges. One of which is to dimensionality reduction.

\begin{figure}[t]
\begin{center}
\includegraphics[width=1\linewidth]{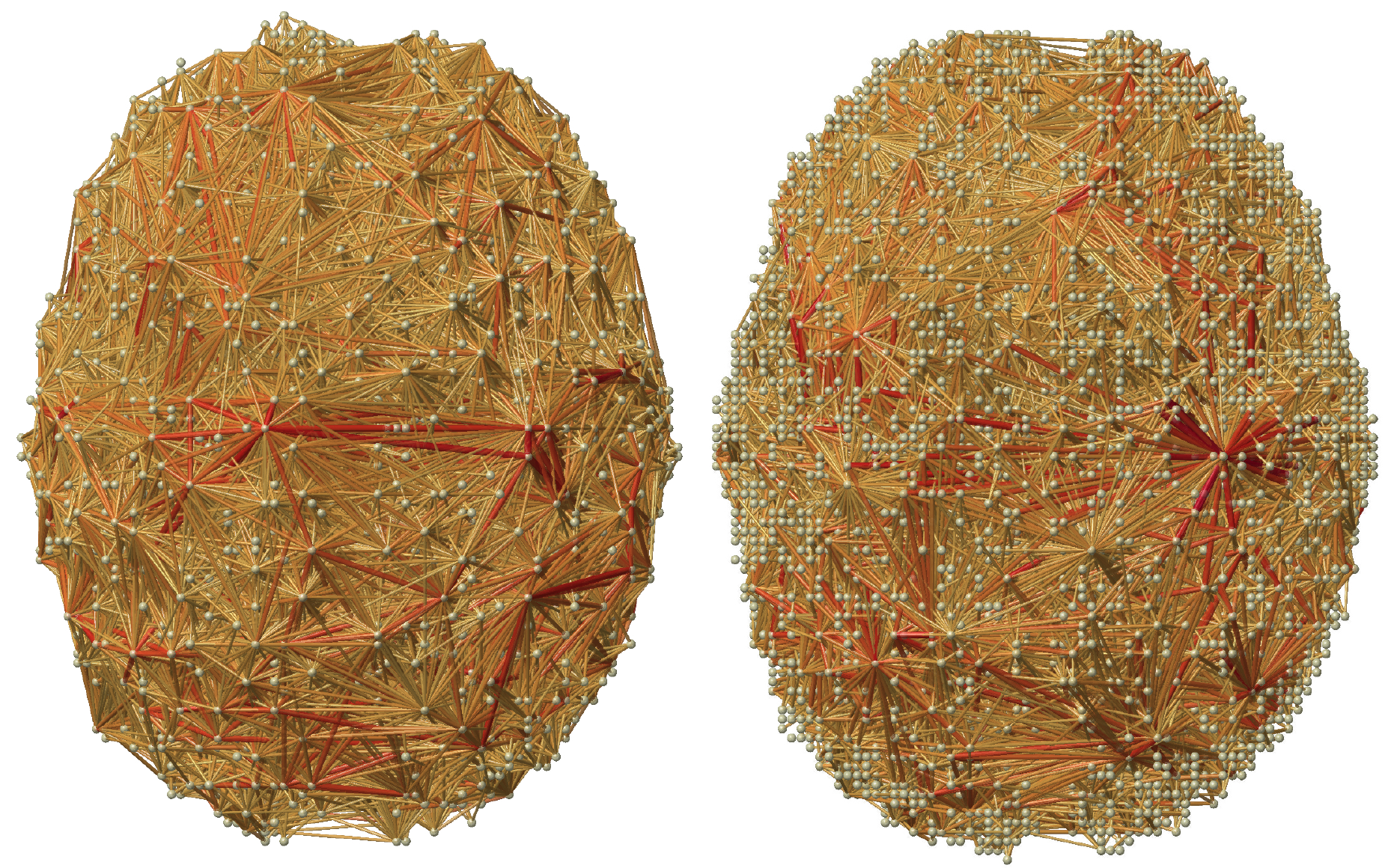}
\caption{rs-fMRI correlation network of two subjects from HCP with more than 25000 nodes. Identifying cycles and computing the number of cycles can be computationally demanding in this type of dense correlation network since persistent homology computations are not very scalable. }
\label{fig:persist-cycle}
\end{center}
\end{figure}

Given $n$ scatter points data in a $k$ dimensional space, the amount of information grows exponentially 
as $k^n$. In contrast, persistent homology compresses the data to $k$ functions 
$f_0 (\epsilon),$ $f_1 (\epsilon),$ $\cdots, f_{k-1} (\epsilon)$ of filtration value $\epsilon$. 
Topological features such as the duration of birth to death of $k$-cycles, Betti numbers and the number of such $k$-cycles are used for such functions. If we further bin the filtration parameter into $N_{bin}$ intervals, 
the amount of information characterized by persistent homology grows only linearly with $k$ as $k N_{bin}$. This huge compression of data can facilitate the search for its underlying patterns in big data. Even so, performing TDA on large dataset is daunting. Persistent homology does not scale well with increased network size (Figure \ref{fig:persist-cycle}). The computational complexity of persistent homology grows rapidly with the number of simplices  \citep{topaz.2015}. With $n$ nodes, the size of the $k$-skeleton  grows as $n^{k+1}$. Homology calculations are often done by Gaussian elimination, and if there are $N_{simplicies}$ simplices, it takes ${\cal O} (N_{simplicies}^3)$ time to perform. In a $d$ dimensional embedding space, this computational time can be estimated as ${\cal O}(n^{3k+3})$ \citep{solo.2018}. Thus, a full TDA computation of Rips complex is exponentially costly; it becomes infeasible when the number of nodes or the dimension of the embedding space is large.  This can easily happen when one tries to use brain networks at the voxel level resolution. 
Thus, there have been many algorithm development in computing Rips complex approximately but fast for large-scale data. There are two broad approaches to make the TDA computation scalable: 1) by sampling a subset of the nodes and 2) by constructing alternate filtrations consisting of significantly smaller number of simplices while keeping the set of nodes unchanged that give the same or approximate persistent homologies of the filtration. This section is focused on the detailed review of the first approach. The construction of alternate filtrations such as graph filtrations are reviewed in the next section.

The second approach speeds up the computation by reducing the number of simplices. The alpha filtration based on Delaunay triangulation is an example of an alternate filtration with a provable guarantee that it consists of smaller number of simplicies than the Rips filtration \citep{otter.2017}. The worst-case number of simplices in the alpha filtration in dimension $k=3$ scales as ${\cal O} (n^2)$. The benefits of a sparser representation than Rips filtration cannot be overstated, but the challenge is to construct a simplified filtration without spoiling the stability results of persistent homology. By perturbing the metric of the data space, one can construct a sparsed Rips filtration which has ${\cal O}(n)$ simplicies and can be computed in ${\cal O} (n \log n)$ time \citep{sheehy2013linear}.

\subsection{Witness complex}
\label{sec:witness}

One crucial difference between geometry and topology is that topology can be accurately estimated from a small sample of the full data, while geometry requires more fine-grained information. This reflects the fact that topology is a more basic notion of shape than geometry. This suggests a sampling procedure can significantly reduce the size of a simplicial representation of data while preserving topological information. Since the number of simplices in the Rips complex grows with respect to the number of vertices $N_{\rm vertices}$ as $2^{\mathcal{O}(N_{\rm vertices})}$, being able to evaluate topology on a small sample of the vertices can greatly speed up the computation. A principled procedure for sampling data points to identify the data's topology is given
by \emph{witness complexes} \citep{deSilva.2004}. The idea is to use a small subset of the given point cloud as landmarks, which form the vertices of the complexes. The remaining non-landmark  points are used as witnesses to the existence of simplices spanned by combinations of landmark points.

Let $Z$ be the full point cloud, $L \subset Z$ denote the set of landmark points, and $W \subset Z$ denotes the set of witnesses. $W$ is a dense sample of $Z$. Then $w \in W$ is a {\em weak witness} of a simplex $\sigma = [v_0, v_1, \cdots, v_k] \subset L$ if for all $v \in \sigma$ and  for all $v^{\prime} \in L \backslash \sigma$, $d(w,v) \leq d(w,v')$ \citep{deSilva.2004}. The largest distance from a witness to the simplex should be smaller than the distance to all other points excluding the simplex. A $k$-simplex is witnessed by $w$ if it consists of $w$'s $(k+1)$-th nearest neighbors in $Z$. Let $d_i (w, \sigma)$ is the $(i+1)$-th closest distance from $w$ to one of vertices in $\sigma$. Note $d_k (w, \sigma) = \max_k d(w, v_k)$. Similarly define $d_k(w,L)$ to be the distance from $w$ to its $(k+1)$-th closest landmark point in $L$. Starting with initial random point $v_0$, the witness complex can be defined by induction. For vertices $v_i \in L$, we include the $k$-simplex $\sigma$ in the witness complex ${\rm Wit} (L,W,\epsilon)$ at scale $\epsilon$ if all of its faces belong to ${\rm Wit} (L,W,\epsilon)$ and there exists a witness point $w\in W$ such that
\bqn
\label{eqn:witness}
	  d_k(w,\sigma) \leq \epsilon+d_k(w,L) 
\eqn
for filtration value $\epsilon$ which is analogous to the filtration values in the Rips filtration. The witness complex is the largest simplicial complex with vertices in $L$ whose faces are witnessed by points in $W$ \citep{guibas.2008}.

\begin{figure}
\centering
\includegraphics[width=0.35\textwidth]{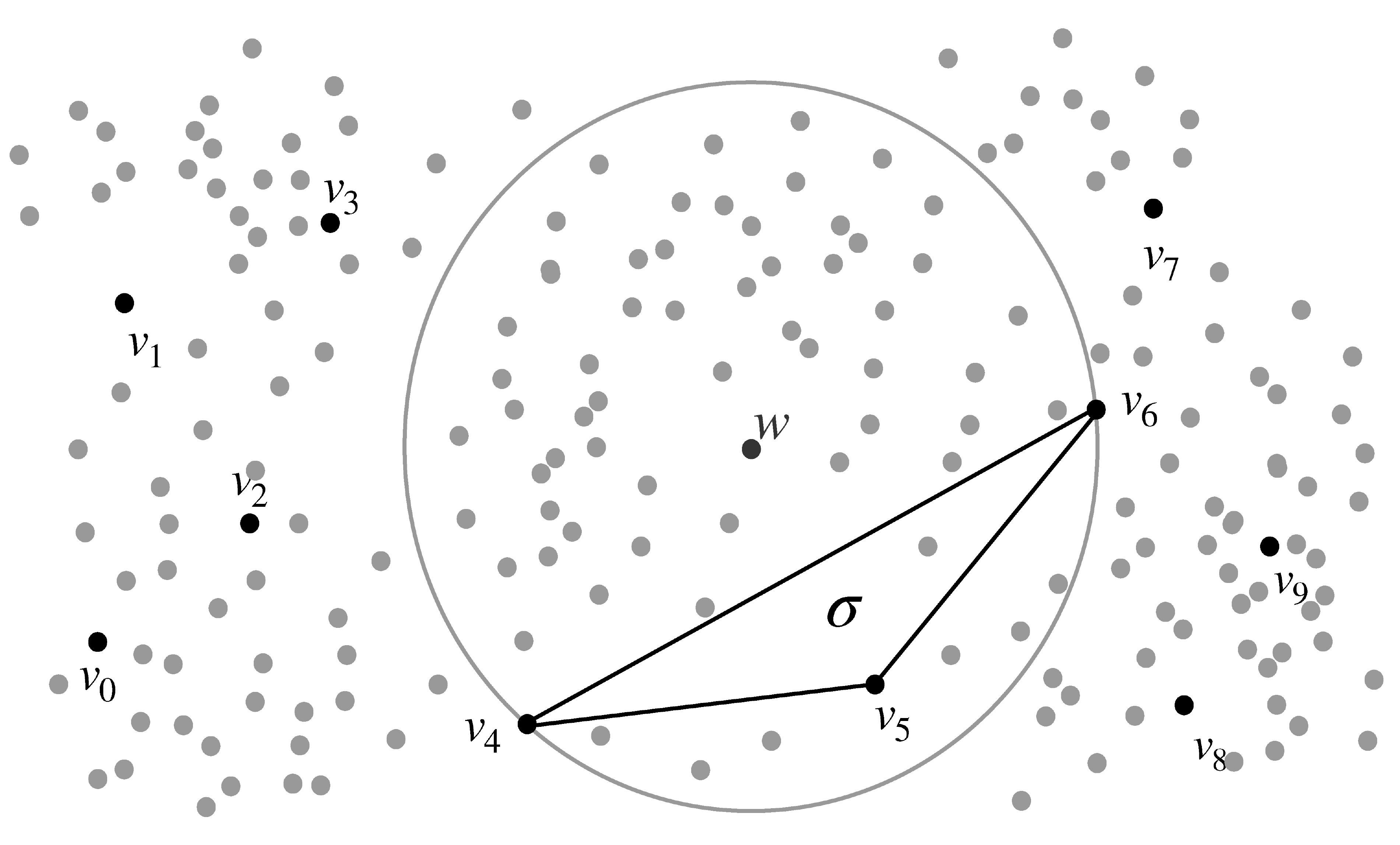}
\caption{In the witness complex, edges and higher-dimensional simplices are built out of vertices in the landmark set $L = \{v_0,v_1, \dots, v_9\}$ and witnessed by other points according to inequality \eqref{eqn:witness}. Point $w$ is a weak witness of the 2-simplex $[v_4, v_5, v_6]$ at filtration scale $\epsilon=0$. This 2-simplex belongs to the witness complex ${\rm Wit} (L,W,0)$ since $\sigma = [v_4,v_5,v_6]$ satisfies  \eqref{eqn:witness} for $k=2$ and all its faces (edges) $[v_4, v_5], [v_5,v_6], [v_6,v_4]$ satisfy \eqref{eqn:witness} for $k=1$.}\label{fig:witness}
\end{figure}

The witness complex depends on the initial choice of landmarks. We want to minimize the number of landmark points to reduce computational time, but yet this landmark set should be representative of the data, as the four points in representing the circle in Figure  \ref{fig:witness}. The landmarks are chosen either randomly or through the \emph{maxmin} algorithm. which tend to select evenly spaced landmarks  \citep{deSilva.2004,cole.2019}. In brain imaging applications, the landmark can be chosen biologically. In building parcellation-based networks, center of individual parcellation can be chosen as landmarks and approximate the overall topology of large-scale brain networks constructed at the voxel-level. The size of a witness complex is determined by the size of the landmark set $L$. Since $L$ represents only a fraction of the points in given point cloud $Z$, witness complexes are much smaller than the Rips complexes corresponding to $Z$. The upper bound $|Z|/|L|\leq 20$ was suggested for 2D surface data \citep{deSilva.2004}. Taking $|Z|/|L|=20$ and denoting by $N_W,N_R$ the sizes of the corresponding witness complex and Rips complex respectively, the worst-case scaling of the number of simplices in the witness complex is \citep{otter.2017,arafat.2019}
$$N_{W}\sim 2^{|L|}\sim 2^{|Z|/20}\sim N_{R}^{1/20}.$$ 
Therefore the witness complex presents a huge computational advantage for large data sets.

Witness complexes can be thought of as approximations of the Delaunay triangulation in 2D \citep{deSilva.2004, attali.2007,guibas.2008,boissonnat.2009}. The advantage of the witness complex is that the ambient dimension does not affect the complexity of the algorithm, while the Delaunay triangulation suffers from the curse of dimensionality.  An algorithm was proposed in \citet{boissonnat.2009} to enrich the set of witnesses and landmarks to preserve the relation between the Delaunay triangulation and the witness complex in $k>2$ dimensions so as to ensure that the witness complex is a good approximation of the underlying topology of data.

Witness complexes were applied in the study of natural images and identifying them with the topology of the Klein bottle \citep{carlsson.2008.JCV} and the study the topology of activity patterns in the primary visual cortex \citep{singh.2008}. Interestingly, spontaneous activation patterns and activity stimulated by natural images were found to have compatible topological structure, that of a two-sphere.
\citep{dabaghian.2012} used witness complexes to model activity in the hippocampus as storing topological information about the local environment.

\subsection{$\alpha$-filtrations}\label{sec:alpha}

The $\alpha$-filtration \citep{edelsbrunner1994three} takes its inspiration from often used Delaunay triangulations, which is a simplicial complex.  The Delaunay triangulation or tetrahedralization has been widely used in brain imaging field for building surface or volumetric meshes from MRI \citep{si.2015,wang.2017.yalin}. 

\begin{figure}[t]
    \centering
    \includegraphics[width=1\linewidth]{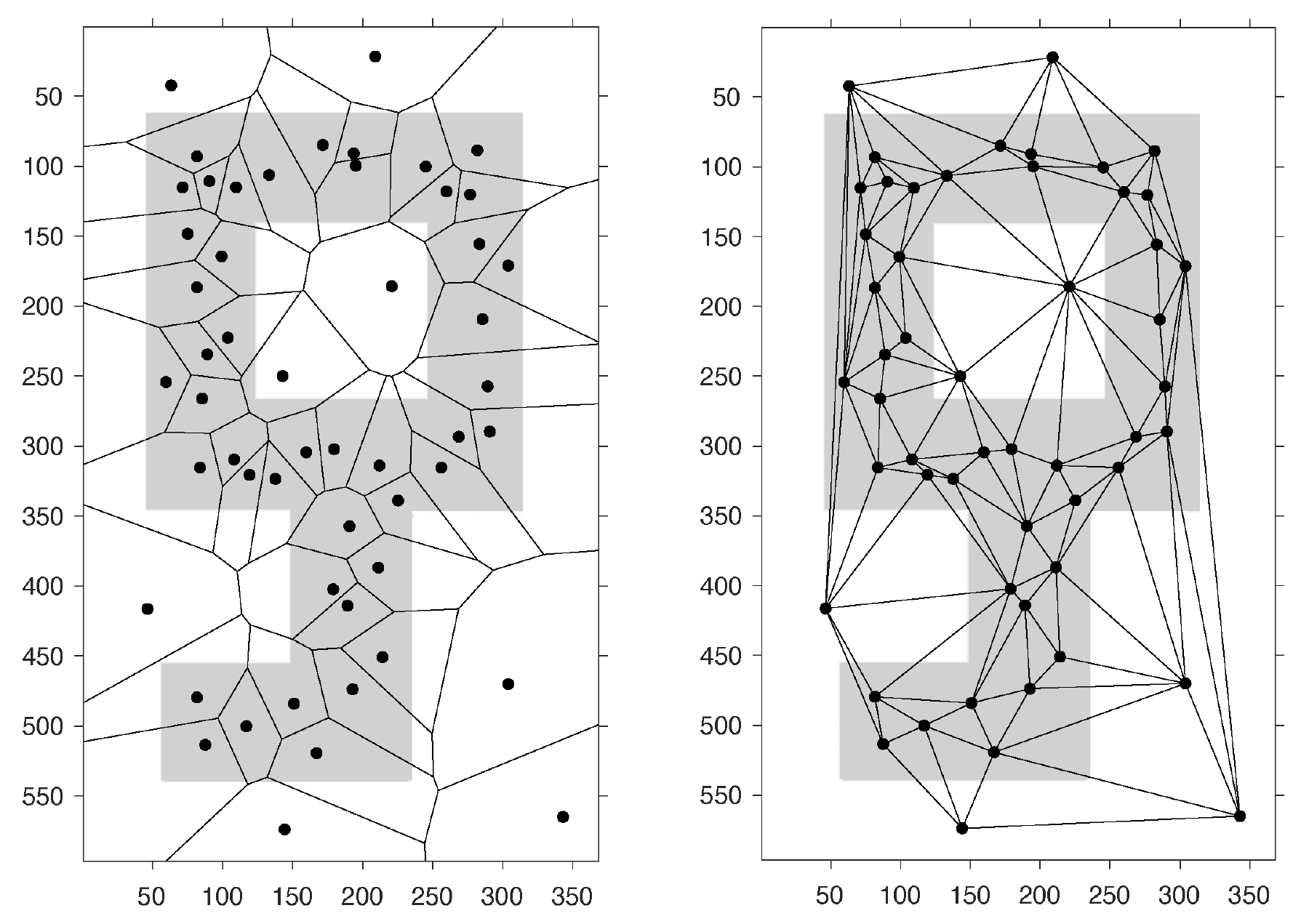}
    \caption{Left: The Voronoi diagram obtained from the point cloud that samples the key shaped object.
    Right: The Delaunay complex, which is the dual graph of the Voronoi diagram.}
    \label{fig:Delaunay}
\end{figure}

Given a set of points $X = \{ x_1, \cdots, x_p \} \subset \mathbb{R}^D$, the \emph{Voronoi cell} around $x_k$ is given by $$V_{k}=\{y\in \mathbb{R}^D ~|~ d(x,x_k) \leq d(x,x_j) \mbox{ for all }  j \neq k\}.$$ 
 The Voronoi cell of point $x_k$ includes all points for which $x_k$ is the nearest point among all points in $X$. The Voronoi cell is a convex set. The collection of Voronoi cells triangulates the domain $\mathbb{R}^D$ and constitutes the \emph{Voronoi diagram} of $X$. The \emph{Delaunay complex} is defined via the Voronoi diagram as $\cup_k V_k$.
An example of this geometric realization is shown in Figure \ref{fig:Delaunay}.  
The Delaunay triangulation is then defined as the dual graph of the Voronoi diagram. The minimum spanning tree of the complete graph consisting of node set $X$ and edge weights $d(x_k,x_j)$  is a subset of the Delaunay triangulation.

The size of the Delaunay complex grows with the number of vertices $N_{\rm vertices}$ as $N_{\rm vertices} \log N_{\rm vertices}$ for $D \leq 2$ and $N_{\rm vertices}^{D/2}$ for $D \geq 3$ \citep{otter.2017}.
This is a clear advantage of the Delaunay complex over the Rips and \v{C}ech complexes whose size grow exponentially with the number of vertices, though the Delaunay complex still suffers from the curse of dimensionality.
Developing efficient algorithms for handling higher dimensional Delaunay complex is a subject of ongoing research \citep{boissonnat2009incremental}. \citet{wang.2017.CNI} used the Delaunay triangulation in building a complex from EEG channel locations.

To further speed up the computation, $\alpha$-complex is often used over Rips complex. The $\alpha$-complex, which is a subcomplex of the Delaunay complex, is defined as follows. For each point $x \in X$ and $\epsilon>0$, let $B_x(\epsilon)$ be the closed ball with center $x$ and radius $\epsilon$. The union of these balls $\cup_{x \in X} B_x (\epsilon)$  is the set of points in $\mathbb{R}^D$ at distance at most $\epsilon$ from at least one of the points of $X$. We then intersect each ball with the corresponding Voronoi cell, $R_x (\epsilon) = B_x (\epsilon) \cap V_x$.
The collection of these sets $\{ R_x (\epsilon) ~|~ x \in X \}$ forms a cover of $\cup_{x \in X} B_x (\epsilon)$, and the $\alpha$-complex of $X$ at scale $\epsilon$ is defined as
\bq
{\rm Alpha} (X, \epsilon) = \{ \sigma \subset X ~|~ \cap_{x \in \sigma} R_{x} (\epsilon) \not=\emptyset \}.
\eq
Since balls and Voronoi cells are convex, the $R_x (\epsilon)$ are also convex. ${\rm Alpha} (X,\epsilon)$ has the same homology as 
$\cup_{x \in X} B_x (\epsilon)$. Not only is ${\rm Alpha} (X,\epsilon)$ a subcomplex of the Delaunay complex, it is also a subcomplex of the \v{C}ech complex.

One potentially undesirable aspect of the $\alpha$-filtration is that the topological objects it identifies die at smaller scales than one would expect when there are outliers. Thus, the $\alpha$-filtration is sensitive to outliers. We can try to remedy this situation by modifying the $\alpha$-filtration to a weighted $\alpha$-filtration that allows balls of different sizes, and building the filtration in a spatially adaptive fashion. 
In applications, this weighted $\alpha$-filtration is often motivated by the underlying physical and biological models.
For example, in studying the persistent homology of large-scale structure data in cosmology \citep{Biagetti:2020skr},  the gravitational potential depends on the mass density of clusters ($0$-simplices) suggesting a varying ball sizes.
In modeling of biomolecules, such as proteins, RNA, and DNA, each atom is represented by a ball whose radius reflects the range of its van der Waals interactions and thus depends on the atom type. For brain network applications, we can adjust the size of balls to follow the contour of either gray matter or white matter of the brain and adoptively build the filtration.

A different modification to the $\alpha$-filtration can be done that reduces the effect of outliers \citep{Biagetti:2020skr}.  In $\alpha$DTM-filtration, we assign the simplices a {\em filtration time} based on the distance-to-measure (DTM) function. Given a set of point $X$, the empirical DTM function is given by 
\begin{equation}
    {\rm DTM}(x)=\frac{1}{k}\left(\sum_{x_i\in N_k(x)}||x-x_i||^p\right)^{1/p}
\end{equation}
where $N_k(x)$ is the list of the $k$ nearest neighbors in $X$ to $x$ \citep{chazal2011geometric}. Often $p=2$ is used. One can build intuition by considering the case $k=1$, in which the DTM function gives the distance to the nearest point in $X$. Increasing $k$ ``smooths'' this distance function so that it takes small values where it is near many points and large values near outliers. Then performing a sublevel filtration using the DTM function, outliers will be added relatively late in the filtration, leading to a smaller effect on the persistent homology.

For large-scale images, DTM function can be evaluated on an image grid  \citep{Xu:2018xnz}. Evaluating on a grid with high resolution and sufficiently large $k$ involves prohibitive computational expense. One way to get around this is to evaluate the DTM function at certain relevant points of the space. This amounts to choosing an efficient triangulation of the ambient space. In \citet{Biagetti:2020skr}, the DTM function was evaluated on the Delaunay complex. This was shown to keep the computational cost low while maintaining the desirable feature of $\alpha$DTM-filtration in reducing the adverse effects of outliers.

\subsection{Combinatorial Morse theory}
Another approach to significantly reduce the number of simplices is Morse filtration. 
Using the combinatorial Morse theory \citep{forman1998morse}, one can reduce the initial complex using geometric and combinatorial methods before performing homology calculations. Combinatorial Morse theory extends the notion of Morse homology for smooth manifolds to discrete datasets. The gradient flow for a smooth manifold is replaced by a partial pairing of cells in a complex. As a result, the Morse complex has thus a significantly smaller size than the original complex, while preserving all homological information. As demonstrated in \citet{harker2010efficiency}, for many complexes the resulting Morse complex is many orders of magnitude smaller than the original.
An efficient preprocessing algorithm was developed in \citet{mischaikow2013morse} that extends combinatorial Morse theory from complexes to filtrations. This proprocessing algorithm maps any filtration
into a Morse filtration with the same persistent homologies, thus significantly reducing both the computational time and the required memory for running the persistence algorithm.

\subsection{Hypergraphs}

The computational overhead in TDA lies in the homology calculations of higher-dimensional simplices. The primary algorithms in persistent homology  requires Gaussian elimination with cubic runtime in the number of simplices \citep{edelsbrunner.2010}. This can cause an additional computational bottleneck when the number of simplices up to dimension $k$ in the Rips filtration is the order of $\mathcal{O}(n^{k+1})$  for $n$ data points \citep{sheehy2013linear}. Thus, it is worthwhile to explore if the run time for analyzing data structure is significantly reduced if we represent the nodes and their relations in terms of \emph{hypergraphs} \citep{torres2020and}. A hypergraph is a generalization of a graph which contains vertex set $V$ and hyperedge set $E$ consisting of vertices allowing polyadic relations between an arbitrary number of vertices.

\begin{figure}[t]
    \centering
    \includegraphics[width = .45\textwidth]{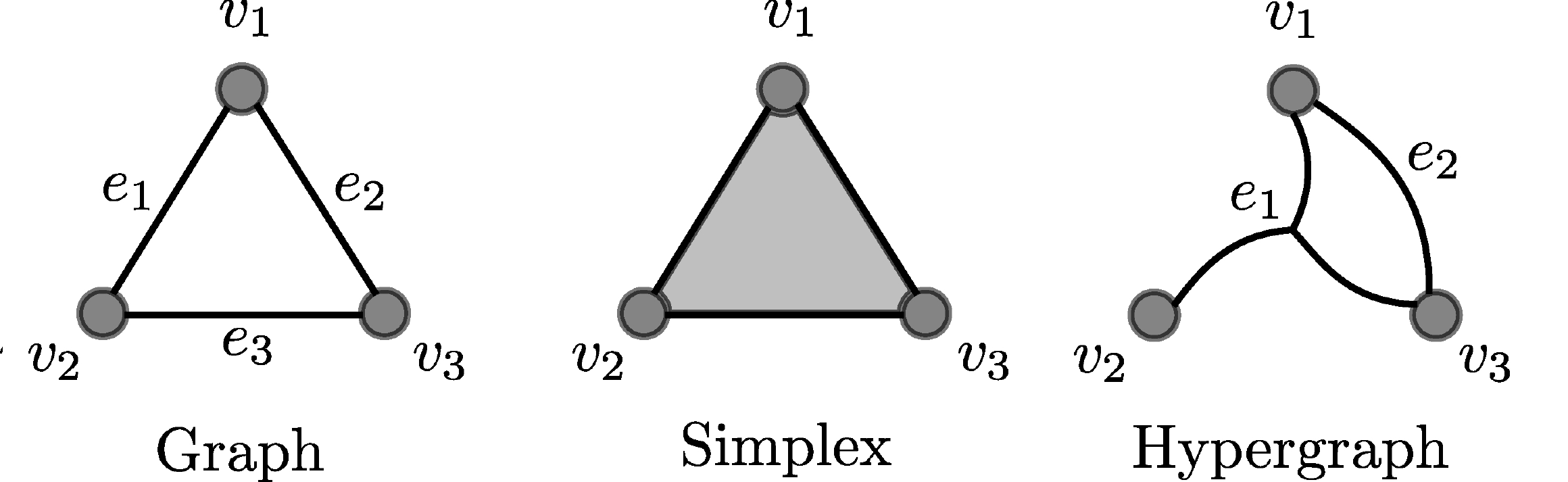}
    \caption{Schematic of graph and hypergraph with three nodes $v_1,v_2,v_3$. The edge set for the graph is  $e_1 =  [v_1,v_2]$, $e_2 = [v_1,v_3 ]$, $e_3 = [v_2,v_3 ]$. The edge set for the hypergraph is given by $e_1 =[ v_1,v_2,v_3 ]$, $e_2 = [v_1,v_3 ]$. A hypergraph allows edges that contain an arbitrary number of vertices, while edges in a graph is limited to only two vertices, preventing it from capturing higher level interactions between vertices. The connection $e_1$ in the hypergraph captures the high level interaction between the three vertices. 
 A graphs on the other hand can only approximate the higher order interaction with a series of pairwise interactions. The simplex captures the high level interaction but also assumes pairwise interactions between all vertices. The hypergraph captures the high level interaction directly.}
    \label{fig:hypergraph}
\end{figure}

This is in contrast to the normally defined graph which can only form connections between two vertices. 
Unlike an edge in a graph or a simplicial complex, a hyperedge can connect any arbitrary number of vertices. Thus, a hypergraph representation of data consists of only 0D and 1D objects. Figure \ref{fig:hypergraph}  displays a hypergraph with three nodes $v_1, v_2, v_3$ representing interactions between all three nodes with hyperedge $e_1$.

We can represent the hypergraph using the incidence matrices or the corresponding adjacency matrices. The incidence matrix $H=(h_{ij})$ is a matrix with rows representing vertices $v_i$ and columns representing edges $e_j$ \citep{estrada2005complex} defined as
\bq
	h_{ij} = \begin{cases}
						1 & v_i \in e_j \\
						0 & v_i \notin e_j
					\end{cases}.
\eq
The adjacency matrix $A$ for a given graph or hypergraph is derived from the incidence matrix 
\bq 
   A = H H^{T} - D,
\eq
where $D$ is the diagonal matrix whose diagonal entries are the degrees of each node \citep{estrada2005complex}:
The adjacency matrix of a hypergraph can be used to compute certain properties of hypergraphs such as centrality and clustering coefficients \citep{estrada2005complex}. The incidence matrix $H$ and adjacency matrix $A$ of the graph in Figure \ref{fig:hypergraph} is given by

\bq
H=
\begin{array}{c}
v_1\\
v_2\\
v_3
\end{array}
\begin{array}{c}
\begin{array}{ccc}
e_{1} & e_{2} & e_{3}\\
\end{array}\\
\left(
\begin{array}{ccc}
 1 & 1 &  0 \\
 1 &  0 & 1\\
 0 &  1 & 1
\end{array}
\right)
\end{array},
\quad
A=
\begin{array}{c}
v_1\\
v_2\\
v_3
\end{array}
\begin{array}{c}
\begin{array}{ccc}
v_1 & v_2 & v_3\\
\end{array}\\
\left(
\begin{array}{ccc}
 1 & 1 &  0 \\
 1 &  0 & 1\\
 0 &  1 & 1
\end{array}
\right)
\end{array}.
\eq
Similarly, the incidence and adjacency matrix of the hypergraph in Figure \ref{fig:hypergraph} is given by
\bq
H=
\begin{array}{c}
v_1\\
v_2\\
v_3
\end{array}
\begin{array}{c}
\begin{array}{cc}
e_{1} & e_{2}\\
\end{array}\\
\left(
\begin{array}{ccc}
 1 & 1 \\
 1 &  0 \\
 1 &  1
\end{array}
\right)
\end{array},
\quad
A=
\begin{array}{c}
v_1\\
v_2\\
v_3
\end{array}
\begin{array}{c}
\begin{array}{ccc}
v_1 & v_2 & v_3\\
\end{array}\\
\left(
\begin{array}{ccc}
 0 & 1 &  2 \\
 1 &  0 & 1\\
 2 &  1 & 0
\end{array}
\right)
\end{array}.
\eq

Simplicial complexes are also capable of capturing higher level interactions, but require representation of higher order relationships using high-dimensional simplices and have the restriction that these relationships have a property known as the \emph{downward inclusion} \citep{torres2020and}. The downward inclusion requires that any subset of $(k+1)$-vertices that form a $k$-simplex must also form a simplex. This reduces the ability of the simplicial complex to represent abstract relationships, such as those used to understand the cross link connectivity of functional brain networks over time \citep{bassett2014cross}, as it assumes the presence of lower level connections via downward inclusion. It is possible to have a high order interaction between three regions of the brain without significant  pairwise interaction between any two regions. Thus, the simplicial complex
based brain network analysis may over enforce additional low hierarchical dependency than needed.

In Figure \ref{fig:hypergraph}, we illustrate the difference between a graph, a simplex, and a hypergraph in capturing a high level relationship between three vertices. The graph can only capture dyadic relationship between two nodes. For higher order interaction between all three nodes $v_1, v_2, v_3$, additional model is needed. The simplex represents the higher level relationships between all three vertices by the filled-in triangle but requires that all dyadic connections between all node pairs exist, which may not necessarily be true in brain networks. However, the hypergraph captures only the triadic relationship between the vertices without lower level dyadic relationship.

A hypergraph representation of data consists of only 0D (vertices) and 1D (hyperedges) objects making it a low dimensional representation of higher order interactions within a dataset. Hypergraphs provide improved flexibility over simplicial representations used in TDA and accurately capture higher level relationships in complex networks, such as those used to understand covariant structures in neurodevelopment \citep{torres2020and,gu2017functional}. The hypergraph has found many applications in functional brain imaging studies. Hypergraph topology has been used to differentiate brain patterns during attention-demanding tasks, memory-demanding tasks, and resting states \citep{davison2015brain}. Hypergraph topology has also been used in understanding differences in functional connectivity in brains over the human lifespan \citep{davison2016individual}. It has also been used in the diagnosis of brain disease and in the identification of connectome biomarkers \citep{jie2016hyper, zu2016identifying} The use of hypergraphs extends beyond functional networks where they have been used in multi-atlas and multi-modal hippocampus segmentation by developing hypergraphs that represent topological connections across multiple images  over different imaging modalities \citep{dong2015multi}.

Many of the reported methods in literature use simple topological descriptors of hypergraphs, such as number of cross links, hyperedge sizes or clustering coefficients. There has been little application of the concepts of TDA to hypergraphs directly. However, there has been work some preliminary works connecting TDA to hypergraphs including the the analysis of Betti numbers of hypergraphs, the cohomology of hypergraphs, and the embedded homology of hypergraphs \citep{chung1992cohomological,emtander2009betti,bressan2016embedded}.

The application of TDA methods to hypergraphs may yield faster running times as the homology of high dimensional simplices are no longer needed to represent high order interactions, and could result in more meaningful analysis as constraints such as downward inclusion would no longer be required \citep{lloyd2016quantum}.

\section{Persistent homology in brain networks}
For adapting persistent homology to brain network data, {\em graph filtration} was introduced   \citep{lee.2011.MICCAI,lee.2012.TMI}. Instead of analyzing networks at one fixed threshold, we build the collection of nested networks over every possible threshold. The graph filtration is a threshold-free framework for analyzing a family of graphs but requires 
hierarchically building  nested subgraph structures. 
The graph filtration framework shares similarities to existing multi-thresholding or multi-resolution network 
models  that use somewhat arbitrary multiple thresholds or scales  
\citep{achard.2006,he.2008,kim.2015,lee.2012.TMI}. However such approaches are 
mainly exploratory and mainly used to visualize graph feature changes over different thresholds without quantification. Persistent homology, on the other hand,  quantifies such feature changes in a coherent way.

\subsection{Graph filtration}

The graph filtration has been the first type of filtrations applied in brain networks and now considered as the de fact baseline filtrations in the field \citep{lee.2011.MICCAI,lee.2012.TMI}. Euclidean distance is often used metric in building filtrations in persistent homology \citep{edelsbrunner.2010}. Most brain network studies  also use the Euclidean distances for building graph filtrations \citep{lee.2011.ISBI,lee.2012.TMI,chung.2015.TMI,chung.2017.IPMI,petri.2014,khalid.2014,cassidy.2015,wong.2016,anirudh.2016,palande.2017}. Given weighted network $\mathcal{X}=(V, w)$ with edge weight $w = (w_{ij})$, the binary network $\mathcal{X}_{\epsilon} =(V, w_{\epsilon})$ is a graph consisting of the node set $V$ and the binary edge weights 
$w_{\epsilon} =(w_{\epsilon,ij})$ given by 
\bqn w_{\epsilon,ij} =   \begin{cases}
1 &\; \mbox{  if } w_{ij} > \epsilon;\\
0 & \; \mbox{ otherwise}.
\end{cases}
\label{eq:case}
\eqn
Note \citet{lee.2011.MICCAI,lee.2012.TMI} defines the binary graphs by thresholding above such that $w_{\epsilon,ij} =1$ if $w_{ij} < \epsilon$ which is consistent with the definition of the Rips filtration. However, in brain connectivity,  higher value  $w_{ij}$ indicates stronger connectivity so we usually thresholds below  \citep{chung.2015.TMI}.

\begin{figure}[t]
\centering
\includegraphics[width=1\linewidth]{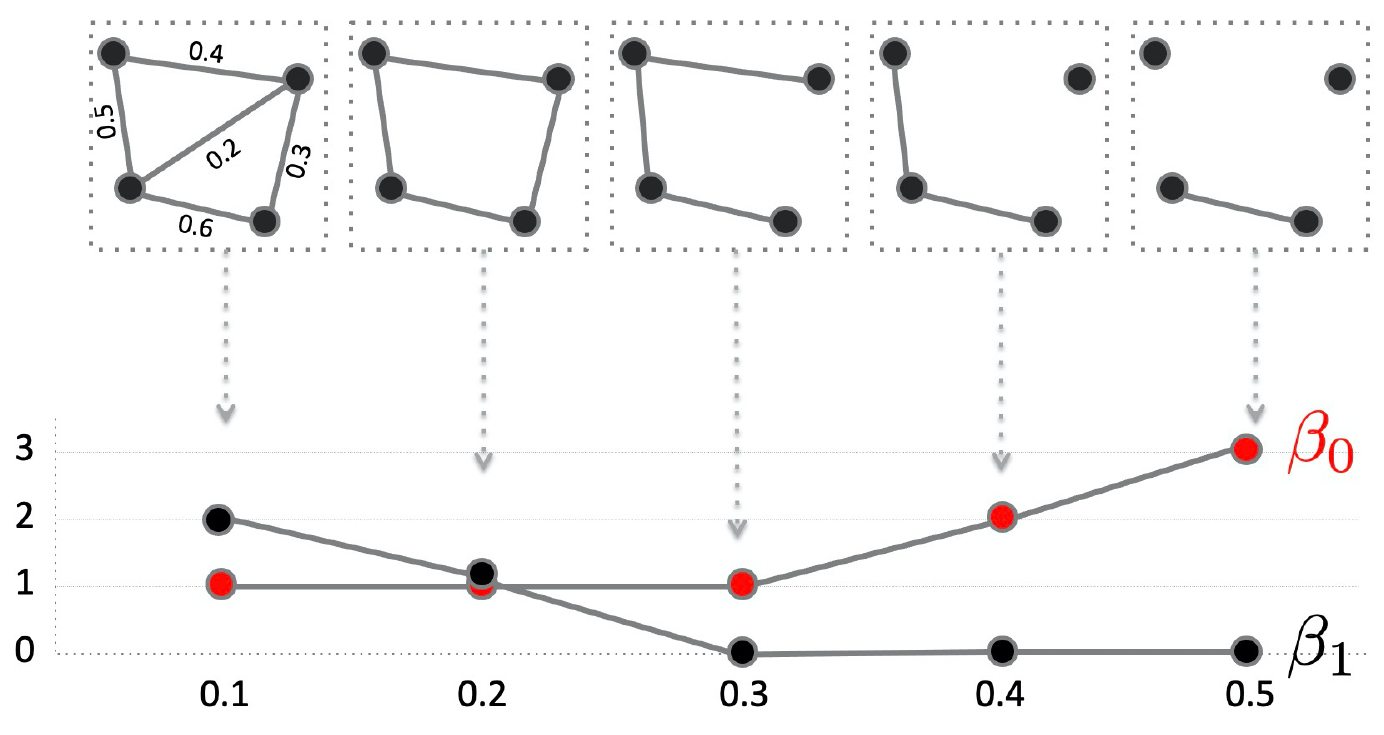}
\caption{Schematic of graph filtration and Betti curves. 
We sort the edge weights in an increasing order. We threshold the graph at filtration values and obtain binary graphs. The thresholding is performed sequentially by increasing the filtration values. The 0-th Betti number $\beta_0$, which counts the number of connected components, and the first Betti number $\beta_1$, which counts the number of cycles, is then plotted over the filtration. The Betti plots curves monotone in graph filtrations.}
\label{fig:persistent-betti01}
\end{figure}

\begin{figure}[t]
\centering
\includegraphics[width=1\linewidth]{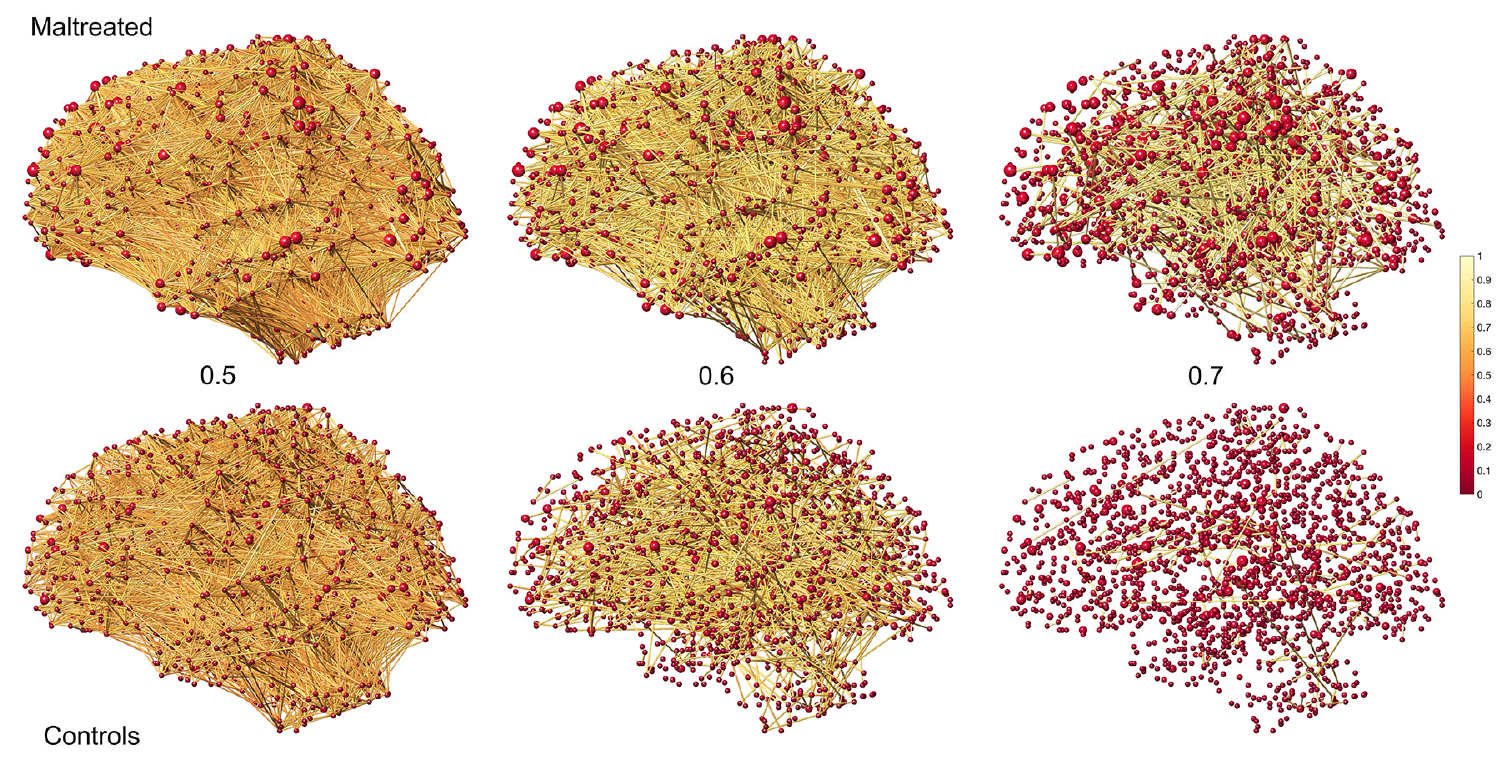}
\caption{Graph filtrations of maltreated children vs. normal control subjects on FA-values \citep{chung.2015.TMI}. The Pearson correlation is used as filtration values at 0.5, 0.6 and 0.7. maltreated subjects show much higher correlation of FA-values indicating more homogeneous and less varied structural covariate relationship.}
\label{fig:GF-maltreated}
\end{figure}

Note $w_{\epsilon}$ is the adjacency matrix of $\mathcal{X}_{\epsilon}$, which is a simplicial complex consisting of $0$-simplices (nodes) and $1$-simplices (edges)  \citep{ghrist.2008}. In the metric space $\mathcal{X}=(V, w)$, the Rips complex $\mathcal{R}_{\epsilon}(X)$ is a simplicial complex whose $(p-1)$-simplices correspond to unordered $p$-tuples of points that satisfy $w_{ij} \leq \epsilon$ in a pairwise fashion \citep{ghrist.2008}. While the binary network $\mathcal{X}_{\epsilon}$ has at most 1-simplices, the Rips complex can have at most $(p-1)$-simplices . Thus,  $\mathcal{X}_{\epsilon} \subset \mathcal{R}_{\epsilon}(\mathcal{X})$ and its compliment $\mathcal{X}_{\epsilon}^c \subset \mathcal{R}_{\epsilon}(\mathcal{X})$. Since a binary network is a special case of the Rips complex, we also have
$$\mathcal{X}_{\epsilon_0}  \supset \mathcal{X}_{\epsilon_1}  \supset \mathcal{X}_{\epsilon_2} \supset \cdots$$
and equivalently 
$$\mathcal{X}_{\epsilon_0}^c  \subset \mathcal{X}_{\epsilon_1}^c  \subset \mathcal{X}_{\epsilon_2}^c \subset \cdots  $$
for
$0=\epsilon_{0} \le \epsilon_{1} \le \epsilon_{2} \cdots.$ The sequence of such nested multiscale graphs  is defined as the {\em graph filtration} \citep{lee.2011.MICCAI,lee.2012.TMI}. Figure \ref{fig:persistent-betti01} illustrates a graph filtration in a 4-nodes example while Figure \ref{fig:GF-maltreated} shows the graph filtration on structural covariates on maltreated children on 116 parcellated brain regions. fv

Note that $\mathcal{X}_0$ is the complete weighted graph while $\mathcal{X}_{\infty}$ is the node set $V$. By increasing the threshold value, we are thresholding at higher connectivity so more edges are removed. Given a weighted graph, there are infinitely many different filtrations. This makes the comparisons between two different graph filtrations difficult. For network $\mathcal{Y}=(V, z)$ with the same node set but with different edge weight $z$, with different filtration values, 
$\lambda_{0} \le \lambda_{1} \le \lambda_{2} \cdots,$
we have
$$\mathcal{Y}_{\lambda_0}  \supset \mathcal{Y}_{\lambda_1}  \supset \mathcal{Y}_{\lambda_2} \supset \cdots.$$
Then how we compare two different graph filtrations $\{ \mathcal{X}_{\epsilon_i} \}$ and $\{ \mathcal{Y}_{\lambda_i} \}$? For different $\epsilon_j$ and $\epsilon_{j+1}$, we can have identical binary graph, i.e., $\mathcal{X}_{\epsilon_j}  = \mathcal{X}_{\epsilon_{j+1}}$. For graph $\mathcal{X}=(V, w)$ with $q$ unique positive edge weights, the maximum number  of unique filtrations  is $q+1$  \citep{chung.2015.TMI}. 

For a graph with $p$ nodes, the maximum number of edges is $(p^2-p)/2$, which is obtained in a complete graph. If we order the edge weights in the increasing order, we have the sorted edge weights:
$$0 = w_{(0)} <  \min_{j,k} w_{jk} = w_{(1)} < w_{(2)} < \cdots < w_{(q)} = \max_{j,k} w_{jk},$$
where $q \leq (p^2-p)/2$.  The subscript $_{( \;)}$ denotes the order statistic. 
For all $\lambda < w_{(1)}$, $\mathcal{X}_{\lambda} = \mathcal{X}_{0}$ is the complete graph of $V$. For all $w_{(r)}  \leq \lambda < w_{(r+1)} \; (r =1, \cdots, q-1)$, $\mathcal{X}_{\lambda} = \mathcal{X}_{w_{(r)}}$. For all $ w_{(q)} \leq \lambda$, $\mathcal{X}_{\lambda}= \mathcal{X}_{\rho_{(q)}} =V$, the vertex set. Hence, the filtration given by
\bq  \mathcal{X}_{0}  \supset  \mathcal{X}_{w_{(1)}}  \supset  \mathcal{X}_{w_{(2)}}  \supset \cdots  \supset  \mathcal{X}_{w_{(q)}}\label{eq:maximal}\eq
is {\em maximal} in a sense that we cannot have any additional filtration $\mathcal{X}_{\epsilon}$ that is not one of the above filtrations. Thus, graph filtrations are usually given at edge weights. 

The condition of having unique edge weights is not restrictive in practice. Assuming edge weights to follow some continuous distribution, the probability of any two edges being equal is zero. 
For discrete distribution, it may be possible to have identical edge weights. Then simply add Gaussian noise or add extremely small increasing sequence of numbers to $q$ number of edges.

\subsection{Monotone Betti curves}

The graph filtration can be quantified using monotonic function $f$ satisfying
\bqn f ( \mathcal{X}_{\epsilon_0} ) \geq f ( \mathcal{X}_{\epsilon_1} )  \geq f ( \mathcal{X}_{\epsilon_2} )  \geq \cdots   \label{eq:B} \eqn
or
\bqn f ( \mathcal{X}_{\epsilon_0} ) \leq f ( \mathcal{X}_{\epsilon_1} )  \leq f ( \mathcal{X}_{\epsilon_2} )  \leq \cdots   \label{eq:B2} \eqn

\begin{figure}[t]
\centering
\includegraphics[width=1\linewidth]{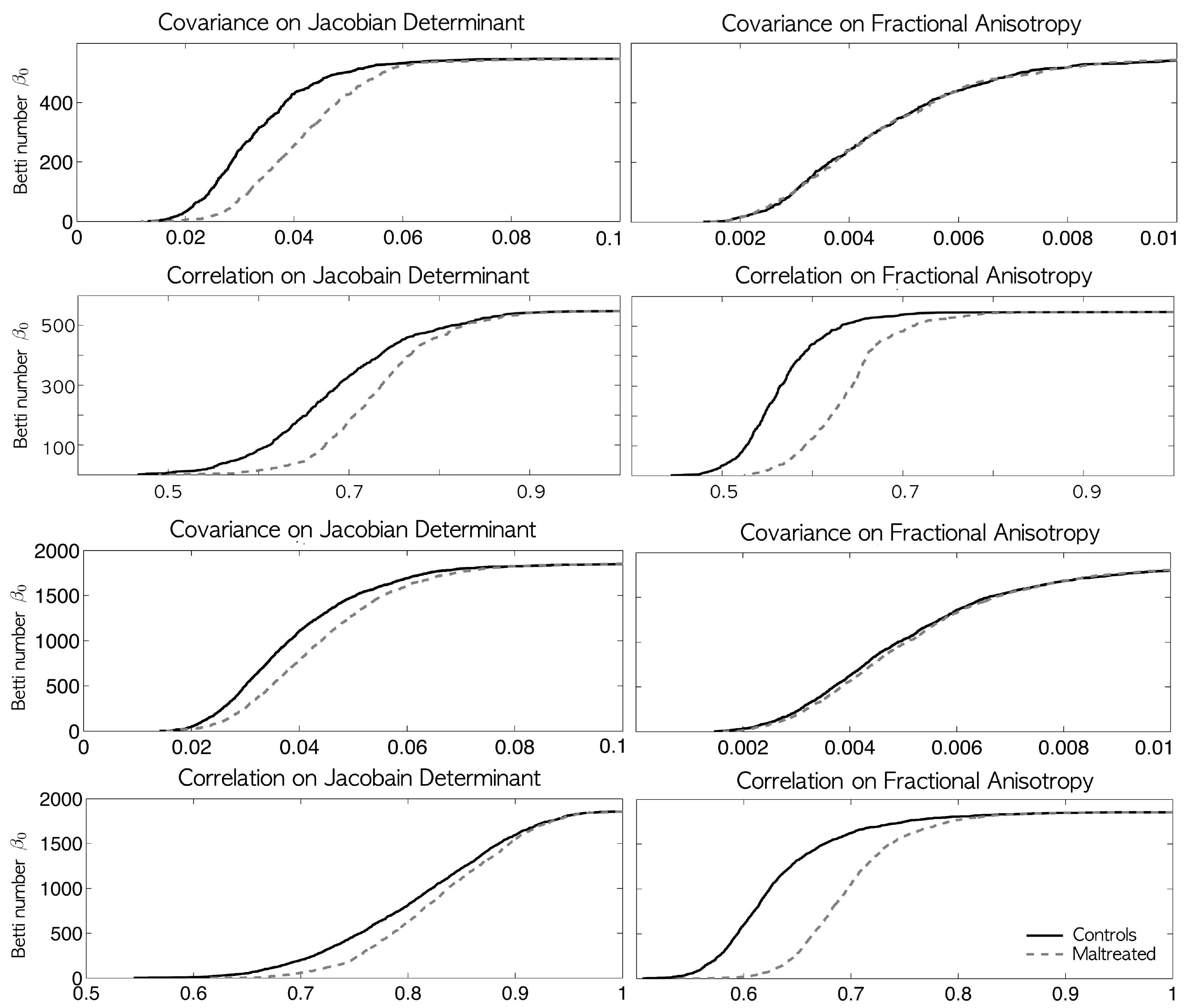}
\caption{The Betti curves on the covariance correlation matrices for Jacobian determinant (left column) and fractional anisotrophy (right column) on 548 (top two rows) and 1856 (bottom two rows) nodes \citep{chung.2015.TMI}. Unlike the covariance, the correlation seems to shows huge group separation between normal and maltreated children visually. However, in all 7 cases except top right (548 nodes covariance for FA), statistically significant differences were detected using the rank-sum test on the areas under the Betti-plots ($p$-value $< 0.001$).  The shapes of Betti-plots are consistent between the studies with different node sizes indicating the robustness of the proposed method over changing number of nodes.}
\label{fig:persist-betti0plots}
\end{figure}

The number of connected components (zeroth Betti number $\beta_0$) and the number of cycles (first Betti number $\beta_1$) satisfy the monotonicity (Figures \ref{fig:persistent-betti01} and \ref{fig:persist-betti0plots}). The size of the largest cluster also satisfies a similar but opposite relation of monotonic increase. There are numerous  monotone graph theory features \citep{chung.2015.TMI,chung.2017.IPMI}.

\begin{figure}[t]
\includegraphics[width=1\linewidth]{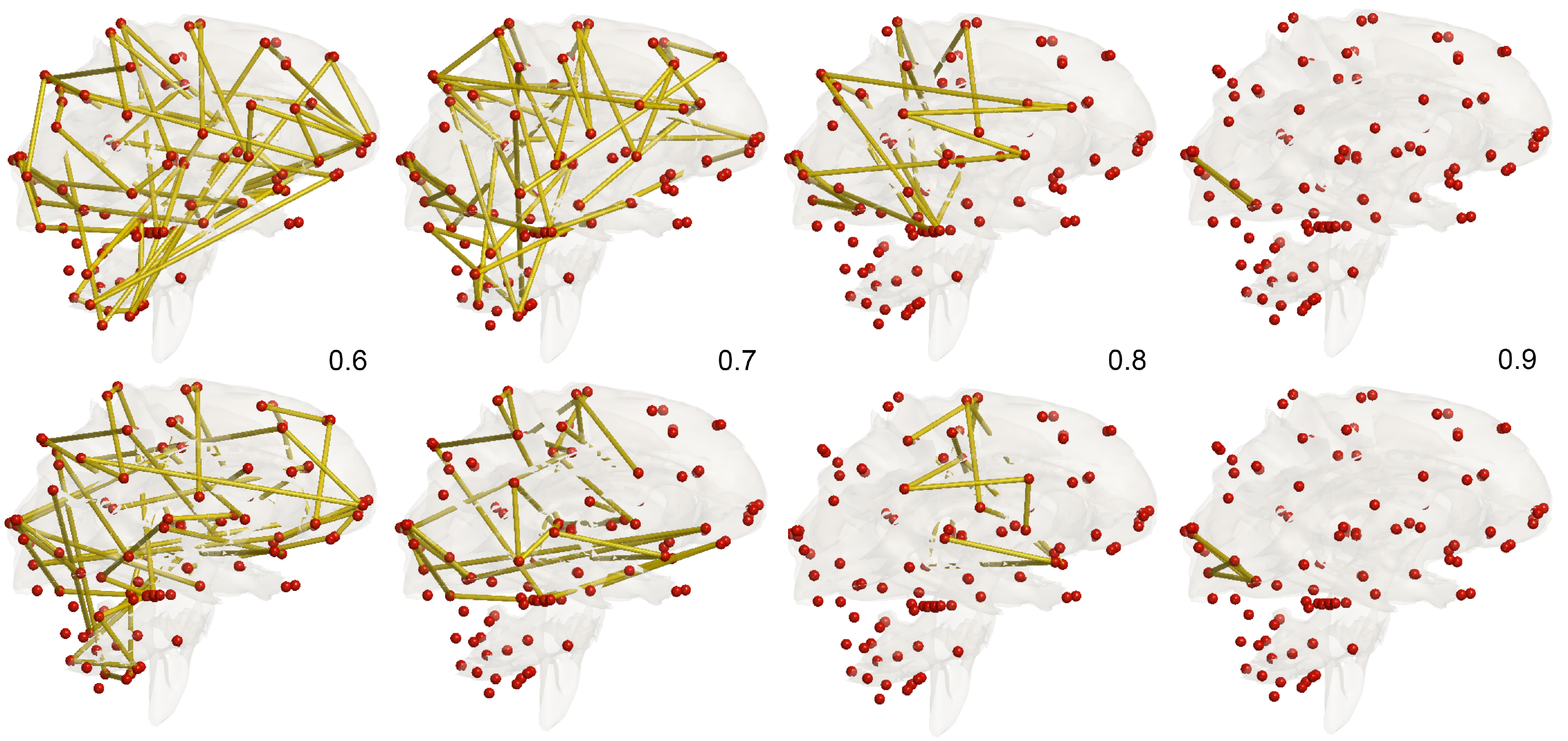}
\caption{The largest cycle at given correlation thresholds on rs-fMRI. Two representative subjects in HCP were used \citep{chung.2019.ISBI}. As the threshold increases, the length of cycles decreases monotonically.}
\label{fig:graphfiltration1}
\end{figure}

For graphs, $\beta_1$ can be computed easily as a function of $\beta_0$. Note that the Euler characteristic $\chi$ can be computed in two different ways
\bq \chi &=& \beta_0 - \beta_1 + \beta_2  - \cdots \\
 &=& \# nodes - \# edges + \# faces - \cdots,
 \eq
 where $\# nodes, \# edges, \# faces$ are the number of nodes, edges and faces. However, graphs do not have filled faces and Betti numbers higher than $\beta_0$ and $\beta_1$ can be ignored. Thus, a graph with $p$ nodes and $q$ edges is given by  \citep{adler.2010}
$$\chi = \beta_0 - \beta_1 = p - q.$$
Thus, $$\beta_1 = p - q - \beta_0.$$
In a graph, Betti numbers $\beta_0$ and $\beta_1$ are monotone over filtration on edge weights \citep{chung.2019.ISBI,chung.2019.NN}. When we do filtration on the maximal filtration in (\ref{eq:maximal}), edges are deleted one at a time.  
Since an edge has only two end points, the deletion of an edge disconnect the graph into at most two. Thus, the number of connected components ($\beta_0$) always increases and the increase is at most by one. Note $p$ is fixed over the filtration but $q$ is decreasing by one while $\beta_0$ increases at most by one. Hence, $\beta_1$ always decreases and the decrease is at most by one. Further, the length of the largest cycles, as measured by the number of nodes, also decreases monotonically (Figure \ref{fig:graphfiltration1}).

Identifying connected components in a network is important to understand in decomposing the network into disjoint subnetworks. The number of connected components (0-th Betti number) of a graph is a topological invariant that measures the number of structurally independent or disjoint subnetworks.  There are many available existing algorithms, which  are not related to persistent homology, for computing the number of connected components including the Dulmage-Mendelsohn decomposition \citep{pothen.1990}, which has been widely used for decomposing sparse matrices into block triangular forms in speeding up matrix operations.

In graph filtrations, the number of cycles increase or decreases as the filtration value increases. The pattern of monotone increase or decrease can visually show how the topology of the graph changes over filtration values. The overall pattern of  {\em Betti curves} can be used as a summary measure of quantifying how the graph changes over increasing edge weights \citep{chung.2013.MICCAI} (Figure \ref{fig:persistent-betti01}). The Betti curves are related to  barcodes. The Betti number is equal to the number of bars in the barcodes at the specific filtration value.

\subsection{Graph filtration in trees}

Binary trees have been a popular data structure to analyze using persistent homology in recent years \citep{bendich.2016,li.2017}. Trees and graphs are 1-skeletons, which are Rips complexes consisting of only nodes and edges. However, trees do not have 1-cycles and can be quantified using up to 0-cycles only, i.e., connected components, and higher order topological features can be simply ignored. However, \citet{garside.2020} used somewhat inefficient filtrations in the 2D plane that increase the radius of circles from the root node or points along the circles. Such filtrations will produces persistent diagrams (PD) that spread points in 2D plane. Further, it may create 1-cycles. Such PD are difficult to analyze since scatter points do not correspond across different PD. For 1-skeleton, the {\em graph filtration} offers more efficient alternative \citep{chung.2019.NN,song.2020.arXiv}.

\begin{figure*}
\centering
\includegraphics[width=1\linewidth]{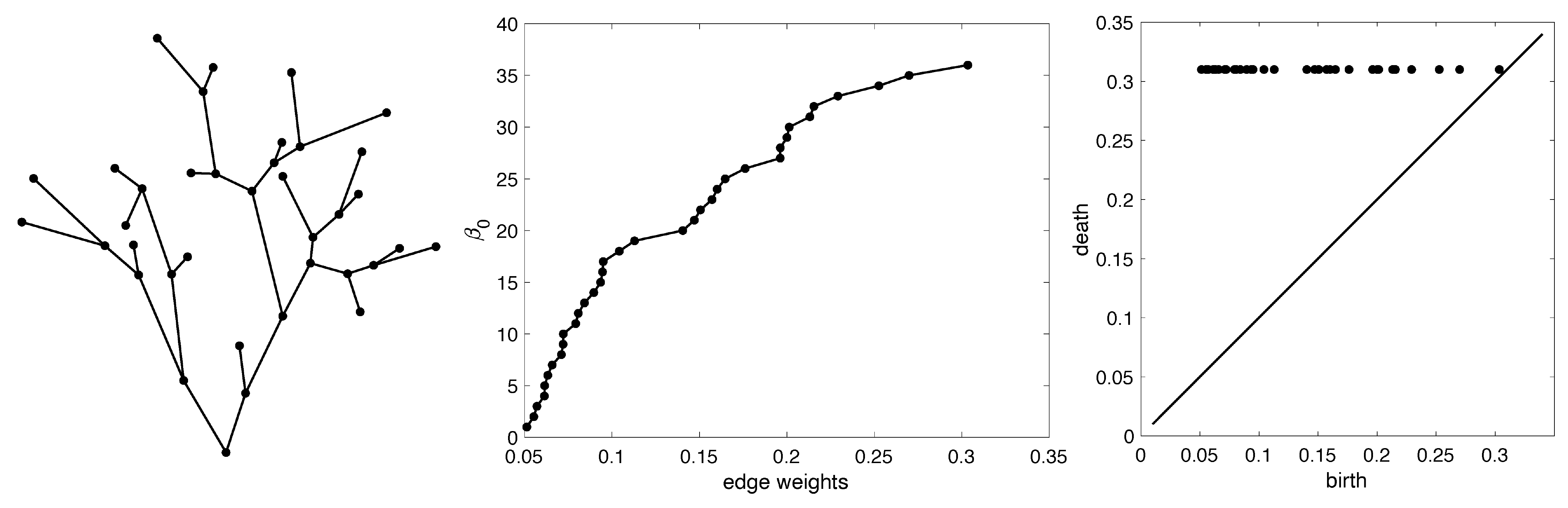}
\caption{Left: binary tree used in  \citet{garside.2020}. Middle: $\beta_0$-curve over graph filtration. Edge weights of the tree is used as the filtration values.  Right: The points in the persistent diagram all lined up at $y=0.31$, which is arbitarily picked to be larger than the maximum edge weight 0.3034.}
\label{fig:filtrationtree}
\end{figure*}

Consider a tree $\mathcal{T}=(V, w)$ with node set $V=\{1, 2, \cdots, p \}$ and weighted adjacency matrix $w$. If we have binary tree with binary adjacency matrix, we add edge weights by taking the distance between nodes $i$ and $j$ as the edge weights $w_{ij}$ and build a weighted tree with $w=(w_{ij})$. For a tree $T$ with $p \geq 2$ nodes and unique $p-1$ positive edge weights $w_{(1)} < w_{(2)} < \cdots < w_{(p-1)}$. Threshold $\mathcal{T}$ at $\epsilon$ and define the binary tree $\mathcal{T}_{\epsilon} =(V, w_{\epsilon})$ with edge weights  $w_{\epsilon} = (w_{\epsilon,ij}), w_{\epsilon,ij} =  1$ if $w_{ij} > \epsilon$ and 0 otherwise. Then we have graph filtration on trees
\bqn  \mathcal{T}_{w_{(0)}}  \supset  \mathcal{T}_{w_{(1)}}  \supset  \mathcal{T}_{w_{(2)}}  \supset \cdots  \supset  \mathcal{T}_{w_{(p-1)}}.\label{eq:maximal2}\eqn
Since all the edge weights are above filtration value $w_{(0)}=0$, all the nodes are connected, i.e., 
$\beta_0(w_{(0)}) = 1$. Since no edge weight is above the threshold  $w_{(q-1)}$, $\beta_0( w_{(p-1)}) = p$. 
Each time we threshold, the tree splits into two and  the number of components increases exactly by one in the tree  \citep{chung.2019.NN}. Thus,
we have 
$$\beta_0(\mathcal{T}_{w_{(1)}}) = 2, \beta_0(\mathcal{T}_{w_{(2)}}) = 3, \cdots, \beta_0(\mathcal{T}_{w_{(p-1)}}) = p.$$ 
Thus, the coordinates for the 0-th Betti curve is given by
$$(0, 1), (w_{(1)}, 2), \cdots,  (w_{(2)}, 3), (w_{(p-1)}, p), (\infty, p).$$

All the 0-cycles (connected components) never die once they are bone over graph filtration. For convenience, we simply let the death value of  0-cycles  at some fixed number $c > w_{(q-1)}$. Then PD of the graph filtration is simply 
$$(w_{(1)},c), (w_{(2)}, c), \cdots, (w_{(q-1)}, c)$$ 
forming 1D scatter points along the horizontal line $y=c$ making various operations and analysis on PD significantly simplified  \citep{song.2020.arXiv}. Figure \ref{fig:filtrationtree} illustrates the graph filtration and corresponding 1D scatter points in PD on the binary tree used in  \citet{garside.2020}. A different graph filtration is also possible by making the edge weight to be the shortest distance from the root node. However, they should carry the identical topological information.

For a general graph, it is not possible to analytically determine the coordinates for its Betti curves. The best we can do is to compute the number of connected components $\beta_0$ numerically using the single linkage dendrogram method (SLD) \citep{lee.2012.TMI}, the Dulmage-Mendelsohn decomposition \citep{pothen.1990,chung.2011.SPIE} or through the Gaussian elimination \citep{deSilva.2007,carlsson.2008,edelsbrunner.2002}.

\subsection{Node-based filtration} Instead of doing graph filtration at the edge level, it is possible to build different kind of filtrations at the node level \citep{hofer.2020,wang.2017.CNI}. Consider graph $G=(V,E)$ with nodes $V={1, 2, \cdots, p}$ and node weights
$w_{i}$ defined at each node $i$. With threshold $\lambda$, define a binary network 
$G_{\lambda} = (V_{\lambda}, E_{\lambda})$
where 
$$V_{\lambda} = \{  i  \in V: w_i \le \lambda \}$$
and 
$$E_{\lambda} = \{ (i,j) \in E: \max(w_{i},w_{j}) \le \lambda\}.$$
Note $E_{\lambda} \subset E$ such that two nodes $i$ and $j$ are connected if $\max(w_{i},w_{j}) \le \lambda$. We include a node from $G$ in $G_{\lambda}$ when the threshold $\lambda$ is above its weight, and we connect two nodes in $G_{\lambda}$ with an edge when $\lambda$ is above the larger weight of any of the two nodes. Then we have the {\em node-based} graph filtration
\begin{equation}
\label{eq: node_gf}
G_{w_{(1)} } \subset G_{w_{(2)} } \subset \cdots \subset G_{w_{(q)}}.
\end{equation} 
The filtration \eqref{eq: node_gf} is not affected by reindexing nodes since the edge weights remain the same regardless of node indexing. Each $G_{w_{(j)}}$ in \eqref{eq: node_gf} consists of clusters of connected nodes; as $\lambda$ increases, clusters appear and later merge with existing clusters. The pattern of changing clusters in \eqref{eq: node_gf} has the following properties.

For $w_{(i)} \leq \lambda<w_{(i+1)}$, $G_{\lambda} = G_{w_{(i)}}$, the filtration \eqref{eq: node_gf} is maximal in the sense that no more $G_{\lambda}$ can be added to \eqref{eq: node_gf}. As $\lambda$ increases from $w_{(i)}$ to $w_{(i+1)}$, only the node $v'_{i+1}$ that corresponds to the weight $w_{(i+1)}$ is added in $V_{w_{(i+1)}}$.

Node-based graph filtration was applied to the Delaunay triangulation of EEG channels in the meditation study in \citet{wang.2017.CNI}.  Node weights are the powers at the EEG channels. At each filtration value $\lambda$, both the nodes and edges with weights less than or equal to $\lambda$ are added. The clusters change as $\lambda$ increases. 

Over the years, other filtrations on graph and networks have been developed including clique filtration \citep{petri.2013,stolz.2017}.

\section{Topological distances}

The topological distances and losses are usually built on top of various algebraic representations of persistent homology such as 
barcodes, PD and graph filtrations. The Gromov-Hausdorff (GH) distance is possibly the most popular distance that is originally used to measure distance between two metric spaces \citep{tuzhilin.2016}. It was later adapted to measure distances in persistent homology, dendrograms \citep{carlsson.2008,carlsson.2010,chazal.2009} 
and brain networks \citep{lee.2011.MICCAI,lee.2012.TMI}.  
The probability distributions of GH-distance is unknown. Thus,  the statistical inference on GH-distance has been done through resampling techniques such as jackknife, bootstraps or permutations \citep{lee.2012.TMI,lee.2017.HBM,chung.2015.TMI}, which often cause computational bottlenecks for large-scale networks. To bypass the computational bottleneck associated with resampling large-scale networks, the {\em Kolmogorov-Smirnov (KS) distance} was introduced in \citep{chung.2012.CNA,chung.2017.IPMI,
lee.2017.HBM}. In this section, we review various topological distances.

\subsection{Bottleneck distance}

This is perhaps the most often used distance in persistent homology but it is rarely useful in applications due to the crude nature of metric. Given two networks $\mathcal{X}^1=(V^1, w^1)$ with $m$ cycles and $\mathcal{X}^2=(V^2, w^2)$ with $n$ cycles,  we construct a filtration. Subsequently, PDs 
$$\mathcal{P} (\mathcal{X}^1) = \left\{ (\xi_{1}^{1},\tau_{1}^{1}), \cdots, (\xi_{m}^{1},\tau_{m}^{1}) \right\}$$ 
and $$\mathcal{P} (\mathcal{X}^2) = \left\{ (\xi_{1}^{2},\tau_{1}^{2}), \cdots, (\xi_{n}^{2},\tau_{n}^{2}) \right\}$$ are obtained through the filtration \citep{lee.2012.TMI,chung.2019.NN}. The bottleneck distance between the networks is defined as the bottleneck distance of the corresponding PDs and bound the Hausdorff distance \citep{cohensteiner.2007,edelsbrunner.2008}:
\bqn
D_{B} \big(\mathcal{P} (\mathcal{X}^1),\mathcal{P} (\mathcal{X}^2) \big) = 
\inf_{\gamma} \sup_{1 \leq i \leq m}  \parallel t_i^1 - \gamma(t_i^1) \parallel_{\infty},
\label{eq:D_B}
\eqn
where $t_i^1 =(\xi_{i}^{1},\tau_{i}^{1})  \in \mathcal{P} (\mathcal{X}^1)$ and $\gamma$ is a bijection from $\mathcal{P} (\mathcal{X}^1)$ to $\mathcal{P} (\mathcal{X}^2)$. The infimum is taken over all possible bijections. If  $t_{j}^{2} = (\xi_{j}^{2},\tau_{j}^{2}) = \gamma(t_{i}^{1})$ for some $i$ and $j$, $L_{\infty}$-norm is given by
$$\parallel t_{i}^{1} - \gamma(t_{i}^{1}) \parallel_{\infty} = \max \big( | \xi_{i}^{1}-\xi_{j}^{2}|,| \tau_{i}^{1}-\tau_{j}^{2}| \big).$$ 
The optimal bijection  $\gamma$ is often determined by the Hungarian algorithm \citep{cohensteiner.2007,edelsbrunner.2008}.
Note (\ref{eq:D_B}) assumes $m=n$ such that the bijection $\gamma$ exists.  If $m \neq n$, there is no one-to-one correspondence between two PDs. Then, additional points should be augmented along the diagonal line to match unmatched pairs. This enables us to match short-lived homology classes to zero persistence \citep{chung.2019.NN,cole.2020}.

If the two networks share the same node set $V^1 = V^2$, with $p$ nodes and  the same number of $q$ unique edge weights. If  the graph filtration is performed on two networks, the number of their 0D and 1D cycles that appear and disappear during the filtration is $p$ and $1-p+q$, respectively \citep{chung.2019.NN}. Thus, their persistence diagrams of 0D and 1D cycles always have the same number of points. 

The well known stability theorem  \citep{cohensteiner.2007} states 
$$D_{B} \big(\mathcal{P} (\mathcal{X}^1),\mathcal{P} (\mathcal{X}^2) \big) \leq \|w^1- w^2\|_{\infty}.$$
The stability theorem is established on Morse functions on a compact manifold but should be true for most of applications. 
Since the infinity norm is a crude metric even in the Euclidean space, such stability theorem does not imply statistical sensitivity or robustness.

\subsection{Wasserstein distance}  

The bottleneck distance is not necessarily a sensitive metric and usually performs poorly compared to other distances \citep{lee.2011.MICCAI,chung.2017.CNI}. A more sensitive distance might be the $q$-Wasserstein distance \citep{edelsbrunner.2010} which is related to recently popular {\em optimal transports}.  $q$-Wasserstein distance distance $D_W$ is defined as
$$D_{W} \big(\mathcal{P} (\mathcal{X}^1),\mathcal{P} (\mathcal{X}^2) \big)  = \Big[ \inf_{\gamma} \sum_{i} \|t_i^1 - \gamma(t_i^1) \|_{\infty}^q \Big]^{1/q},$$
where the infimum is taken over all bijections $\gamma$ between $\mathcal{P} (\mathcal{X}^1)$ and $\mathcal{P} (\mathcal{X}^2)$, with the possibility of agumenting unmatched points to the diagonal \citep{chung.2019.NN}.
It can be shown that the $q$-Wasserstein distance  is bounded by
$$D_{W} \big(\mathcal{P} (\mathcal{X}^1),\mathcal{P} (\mathcal{X}^2) \big)  \leq C \big[ \inf \sum \| f - g \|_{\infty}^q \big]^{1/q}$$
for some constant $C$.
Again since the infinity norm is  a crude metric, the stability statement simply implies the distance are well bounded and behave reasonably well but it does not implies statistical sensitivity.

\subsection{Gromov-Hausdorff distance} 

Gromov-Hausdorff  (GH) distance for brain networks is introduced in \citet{lee.2011.MICCAI,lee.2012.TMI}. GH-distance measures the difference between networks by embedding the network into the ultrametric space that represents hierarchical clustering structure of network \citep{carlsson.2010}. The distance $s_{ij}$ between the closest nodes in the two disjoint connected components ${\bf R}_1$ and ${\bf R}_2$ is called the single linkage distance (SLD), which is defined as 
$$ s_{ij} = \min_{l \in {\bf R}_{1}, k \in {\bf R}_{2}} w_{lk}.$$ 
Every edge connecting a node in ${\bf R}_1$ to a node in ${\bf R}_2$ has the same SLD.
SLD is then used to construct the single linkage matrix (SLM) $S = (s_{ij})$. SLM shows how connected components are merged locally and can be used in constructing a dendrogram. SLM is a {\em ultrametric}
 which is a metric space satisfying the stronger triangle inequality $s_{ij} \le \max (s_{ik},s_{kj})$ \citep{carlsson.2010}.
Thus the dendrogram can be represented as a ultrametric space $\mathcal{D} = (V,S),$ which is again a metric space. 
GH-distance between networks is then defined through GH-distance between corresponding dendrograms. Given two dendrograms $\mathcal{D}^{1}=(V,S^{1})$ and $\mathcal{D}^{2}=(V,S^{2})$ with SLM $S^1 = (s^1_{ij})$ and $S^2 = (s^2_{ij})$,
\bqn D_{GH} (\mathcal{D}^1,\mathcal{D}^2) =  \max_{\forall i,j}  | s^1_{ij} - s^2_{ij} |. \label{eq:D_GH} \eqn

%


GH-distance is the maximum of single linkage distance (SLD) differences \citep{lee.2012.TMI}. SLD between two nodes $i$ and $j$ is the shortest distance between two connected components that contain the nodes. The edges that gives the the maximum SLM is the GH-distance between the two networks. Among all topological distances, only GH-distance uses  the single linkage clustering. All other distances do not use the single linkage clustering \citep{chung.2017.CNI}. For instance, bottleneck distance or KS-distance are not related to single linkage clustering \citep{lee.2012.TMI}.

\begin{figure}[t]
\centering
\includegraphics[width=0.8\linewidth]{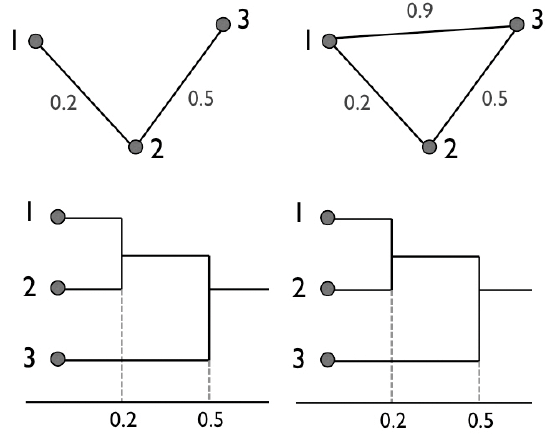}
\caption{Two topologically distinct graphs may have identical dendrograms, which results in zero GH-distance.} 
\label{fig:topology-dendro2}
\end{figure}

The limitation of the SLM is the inability to discriminate a cycle in a graph. Consider two topologically different graphs with three nodes (Figure \ref{fig:topology-dendro2}). However, the corresponding SLM are identically given by
\bq
 \left( \begin{array}{ccc}
0 & 0.2 & 0.5\\ 
0.2 & 0 & 0.5\\
 0.5    &0.5  & 0
\end{array} \right) \mbox{ and }  \left( \begin{array}{ccc}
0 & 0.2 & 0.5\\ 
0.2 & 0 & 0.5\\
 0.5    &0.5  & 0
\end{array} \right).\eq
The lack of uniqueness of SLMs makes GH-distance incapable of discriminating networks with cycles \citep{chung.2012.CNA}.

For the statistical inference on GH-distance, resampling techniques such as jackknife or permutation tests are often used \citep{lee.2011.MICCAI,lee.2012.TMI,lee.2017.HBM}.

\subsection{Kolmogorov-Smirnov distance}

Many TDA distances are not scalable and may not be applicable for large brain networks at the voxel level without simplification on  the underlying algebraic representations. Motivated by the need to build a scalable topological distance, Kolmogorov-Smirnov (KS) distance was introduced in \citet{chung.2013.MICCAI,chung.2015.TMI,chung.2017.IPMI,lee.2017.HBM}. KS-distance is built on top of the graph filtration, a reduced Rips filtration on 1-skeleton. Given two networks $\mathcal{X}^1=(V, w^1)$ and $\mathcal{X}^2=(V, w^2)$, 
KS-distance between $\mathcal{X}^1$ and $\mathcal{X}^2$ is defined as 
$$D_{KS}(\mathcal{X}^1, \mathcal{X}^2) = \sup_{\epsilon \geq 0} \big| \beta_i (\mathcal{X}^1_{\epsilon})  - \beta_i ( \mathcal{X}^2_{\epsilon}) \big|$$
for $\beta_0$ and $\beta_1$ curves on graph filtrations $\mathcal{X}^1_{\epsilon}$ and $\mathcal{X}^2_{\epsilon}$. \
The distance $D_{KS}$ is motivated by the KS-test for determining the equivalence of two cumulative distribution functions  in nonparametric statistics \citep{bohm.2010, chung.2017.IPMI,gibbons.1992}. The distance $D_{KS}$ can be discretely approximated using the finite number of filtrations $\epsilon_1, \cdots, \epsilon_q$:
$$D_q = \sup_{1 \leq j \leq q} \big| \beta_i (\mathcal{X}^1_{\epsilon_j})  - \beta_i ( \mathcal{X}^2_{\epsilon_j}) \big|.$$
If we choose enough number of $q$ such that $\epsilon_j$ are all the sorted edge weights, then  \citep{chung.2017.IPMI}.
$$D_{KS}(\mathcal{X}^1,\mathcal{X}^2) = D_q.$$ 
This is possible since there are only up to $p(p-1)/2$ number of unique edges in a graph with $p$ nodes and the monotone function increases discretely but {\em not continuously}. In practice,  $\epsilon_j$ may be chosen uniformly at discrete intervals.

KS-distance treat the two networks in Figure \ref{fig:topology-dendro2} as identical if  Betti number $\beta_0$ is used as the monotonic feature function. To discriminate cycles, KS-distance on $\beta_1$ should be used. This example illustrates a need for new more sophisticated network distance that can discriminates the higher order topological features beyond connected components.  The advantage of using KS-distance is its easiness to interpret compared to other less intuitive distances from persistent homology. Due to its simplicity, it is possible to determine its probability distribution exactly \citep{chung.2017.IPMI}.

\section{Topological inference and learning}

It is difficult to directly apply existing statistical and machine learning methods to TDA features such as PDs and barcodes that do not have one-to-one correspondence across features. It is difficult to average such features without transformations or smoothing the features first.  Thus, there have been consistent lack of the concept of statistical consistency and statistical methods in the field. One may assume that resampling based statistical inference with minimal statistical assumptions such as the permutation test may work. However, resampling directly on the Rips complex or other derived topological feature may not be feasible for large-scale networks. Thus, there is a strong need to develop a  scalable statistical inference procedures tailored for TDA methods \citep{chung.2017.IPMI,chung.2019.NN}.

\subsection{Statistical consistency} 

Most TDA features have the stability results that show the distance between two topological features are bounded by some known  distance \citep{cohensteiner.2007, adams.2017}. Such stability results do not necessarily provide the {\em statistical consistency} that is required for building proper test statistics  for hypothesis testing that is needed in brain imaging \citep{bubenik.2010}. The test statistic $T_n$, which depends on the sample size $n$, is {\em consistent} if it converge to true value value $T$ in probability as $n$ goes to infinity. For $T_n$ to be consistent, we need
$$\lim_{n \to \infty} P ( | T_n - T| > \epsilon ) =0$$
for all $\epsilon > 0$. 
Most  statistics such as sample mean and $T$-statistics are all consistent. The consistency grantees the convergence of statistical results when enough samples are used.  Existing stability results are mostly on the stability of TDA features  but not about the stability or consistency of the test statistics on such features. So additional investigations are needed to establish the consistency of statistics built on top of TDA features, which were rarely done till now.

\subsection{Exact topological inference} 

The easiest way to build consistent statistical procedures in TDA is using the available topological distances with well established  stability results. Unfortunately, there is no known probability distributions for topological distances except  KS-distance \citep{chung.2019.NN}. The complicated algebraic forms of most TDA distances do not make building probability distributions on them easier. The probability distribution of KS-distance $D_q$ under the null assumption of the equivalence of two Betti curves $\beta_i (\mathcal{X}^1_{\epsilon_j})$ and  $\beta_i ( \mathcal{X}^2_{\epsilon_j})$ is asymptotically given by \citep{chung.2017.IPMI,chung.2019.NN}
\bqn \lim_{q \to \infty}  \Big( D_q /\sqrt{2q} \geq  d  \Big)  = 2 \sum_{i=1}^{\infty} (-1)^{i-1}e^{-2i^2d^2}. \label{eq:pvalue}\eqn
The $p$-value under the null is then computed as
$$\mbox{$p$-value} = 2 e^{-d_{o}^2} - 2e^{-8d_{o}^2} + 2 e^{-18d_{o}^2} \cdots,$$ where the observed value $d_{o}$ is the  least integer greater than $D_{q}/\sqrt{2q}$ in the data. For any large value $d_0 \geq 2$, the second term is in the order of $10^{-14}$ and insignificant. Even for small observed $d_0$, the expansion converges quickly. The KS-distance does not assume any statistical distribution. This is perhaps the only known distribution related to topological distances. The main advantage of the method is that the method avoid using time the consuming permutation test for large-scale networks.

For other distances, resampling techniques such as permutations and bootstraps are needed to determine the statistical distributions.

\subsection{Permutation test}
The permutation test \citep{fisher.1966} is the most widely used nonparametric test procedure in brain imaging, which can be easily adapted for inference on distance and loss functions. It is known as the only {\em exact test} since the distribution of the test statistic under the null hypothesis can be exactly computed by calculating all possible values of the test statistic under every possible combinations. 
Unfortunately, even for modest sample sizes, the total number of permutations is astronomically large and only a small subset of permutations is used in approximating $p$-values. Thus, permutation tests are all approximate in practice. 

The permutation test for two samples are done as follows \citep{chung.2013.MICCAI,efron.1982,lee.2012.TMI}. Two sample test setting is probably the most often encountered  in brain networks. Suppose there are $m$ networks $\mathcal{X}_1, \mathcal{X}_2, \cdots, \mathcal{X}_m$ in Group I and $n$ networks $\mathcal{Y}_1, \mathcal{Y}_2, \cdots, \mathcal{Y}_n$ in Group II. We are interested in testing if networks in two groups differ. Concatenate all the networks and reindex them as 
 $$\mathcal{Z} = (\mathcal{Z}_1, \cdots, \mathcal{Z}_{m+n}) = (\mathcal{X}_1, \cdots, \mathcal{X}_m, \mathcal{Y}_1, \cdots, \mathcal{Y}_n).$$
Following \citet{song.2020.arXiv}, we define the average between-group topological distance as
$$D_B (\mathcal{Z}) = \frac{1}{mn} \sum_{i=1}^m \sum_{j=1}^n D(\mathcal{X}_i, \mathcal{Y}_i).$$
Any topological distance $D$ can be used for this purpose. The smaller distance $D_B$ implies smaller group differences. Thus, $D_B$ can be used as the test statistic for differentiating the groups. For this, we need to determine the empirical distribution of 
 $D_B$ under the null hypothesis of no group difference. Under the null, the group labels are interchangeable and the sample space can be generated by permutations. The permutations are done as follows.  Now permute the indices of $\mathcal{Z}$ in the symmetric group of degree $m+n$, i.e., $S_{m+n}$ \citep{kondor.2007}. The permuted networks are indexed as  $ \mathcal{Z}_{\sigma}$, where $\sigma \in S_{m+n}$ is the permutation. Then we split $ \mathcal{Z}_{\sigma}$ into two parts
$$\mathcal{Z}_{\sigma} = (\mathcal{Z}_{\sigma(1)}, \cdots,  \mathcal{Z}_{\sigma(m)}, \; \mathcal{Z}_{\sigma(m+1)}, \cdots, \mathcal{Z}_{\sigma(m+n)}).$$
Then for each permutation $\sigma$, we compute $D_B (\mathcal{Z}_{\sigma})$. Theoretically, for all $S_{m+n}$, we can compute the the distances. This gives the  empirical estimation of the distribution of $D_B$. The number of permutations exponentially increases and it is impractical to generate every possible permutation. So up to tens of thousands permutations are usually generated to approximate the null distribution. This is an approximate method and a care should be taken to guarantee the convergence. For faster convergence, we can use the transposition test, which iteratively computes the next permutation from the previous permutation in a sequential manner \citep{chung.2019.CNI, song.2020.arXiv}.

\subsection{Learning across brain networks}

Many existing prediction and classification models of brain structures and functions are usually based on a specific brain parcellation scheme. Summary measurements across parcellations are compared through various  standard loss functions such as correlations \citep{huang.2020.NM} and mutual information \citep{thirion.2014}. However, there begin to emerge evidences that brain functions do not correlate well with the fixed parcellation boundaries \citep{rottschy.2012,eickhoff.2016,arslan.2018,kong.2019}. Many existing parcellation schemes give raise to conflicting topological structures over different scales \citep{lee.2012.TMI,chung.2015.TMI,chung.2018.EMBCb}. The topological structure of network at one  parcellation scale may not carry over to different parcellation scales. For instance, the estimated modular structure usually do not have continuity over different resolution parameters or parcellations \citep{betzel.2016,chung.2018.SPL}. Thus, there are needs to develop learning frameworks that provide a consistent and complete picture of  brain anatomical structure and functional processes regardless of the choice of parcellation. One possible solution is
to develop learning framework  across parcellations of different sizes and topology. 

Existing prediction models mainly use regressions, such as general linear models (GLM), logistic regressions or mixed-effect models, that incorporate the accumulated effect of features as the sum of predictive variables in correlating cognitive scores \citep{zhang.2019}. Regression models might be reasonable for determining the group level average patterns. However, the underlying network features that matter most and in what combinations  might be just too complex to be discovered in the regression based predictive models. These challenges can be addressed through  {\em deep learning}, which is methodologically well suited to address all these challenges. 

Due to the popularity of deep learning, there have been recent attempts at trying to incorporate the persistent homology to  deep learning \citep{chen2019topological,hu.2019}. Earlier attempts at incorporating topology into learning framework was not effective since they mainly used topological features into classification frameworks \citep{guo.2018,garin.2019}. Instead of trying to feed the scalar summary of topological features that may not be effective, more efficient approaches seem to use loss and cost functions  that explicitly incorporate topology in learning.

 \subsection{Topological loss}
 
 For various learning tasks such as clustering, classification and regression, traditional losses mostly based on Euclidean distance are often used \citep{chung.2019.NN}. Such loss functions are well suited for measuring similarity and distance between homogenous imaging data that match across subjects. For heterogenous data such as brain sulcal patterns \citep{huang.2020.TMI} or functional activation patterns that may not have voxel-to-voxel correspondence across subjects \citep{desai.2005}, such loss functions are ill-suited or ineffective. Also it is not straightforward to set up loss function between brain parcellations of different sizes and topology. However, the topological distances we reviewed can easily handle  such heterogenous data of different sizes and topology. 
 
In addition to topological distances we reviewed so far, we introduce an additional topological loss function in terms of the barcodes \citep{Edelsbrunner2010,song.2020.ISBI}. The expensive optimization process involved in the Rips filtration takes
 ${\cal O} (d^6)$ run-time for $d$ number of data points 
 \citep{edmonds1972theoretical,kerber2017geometry,Edelsbrunner2010}, making it
impractical in brain networks with far larger number of topological features involving hundreds of connected components and thousands of cycles. Instead of the Rips filtration, if we use  the graph filtration, the optimization problem turns into an optimal matching problem with a significantly reduced run-time of ${\cal O}(d^2 \log d)$ \citep{song.2020.ISBI}.
  
 Consider a network $G = (V,w)$ comprising a set of nodes $V$ and unique positive symmetric edge weights $w = (w_{ij})$. The graph filtration is defined on $G$. Unlike the Rips complex,
there are no more higher dimensional topological features to compute in a graph filtration beyond the 1D topology.
The 0D or 1D barcode corresponding to the network $G$ is a collection  of intervals $[b_i, d_i]$ such that each interval tabulates the life-time of a connected component or a cycle that appears at the first filtration value 
$b_i$ and vanishes at the last filtration value $d_i$.

The number of connected components $\beta_0$ is non-decreasing as $\epsilon$ increases, and so we can simply
represent the 0D barcode of the network using only the set of increasing birth values:
\bq
I_0 (G): \quad b_0 < b_1 < \cdots < b_{m_0}
\eq
Similarly the number of 1-cycles is monotonically decreasing, and thus we can represent the 1D barcode using only the set of increasing death values:
\bq
I_1 (G): \quad d_0 < d_1 < \cdots < d_{m_1}
\eq
Increasing the filtration $\epsilon$ can result in either the birth of a connected component or the death of a cycle but not both at the same time. Therefore, the set of 0D birth values $I_0(G)$ and 1D death values $I_1(G)$ partition the edge weight set $W$ such that $W = I_0 (G) \cup I_1(G)$ with $I_0 (G) \cap l_1(G) =\emptyset$ \citep{song.2020.ISBI}. This decomposition can be effectively used in computing the topological loss efficiently. 

 A topological loss function which measures the topological similarity between two networks can be defined in terms of the 0D and 1D barcodes. 
 Let $G_1 = (V, w^1)$ and $G_2 = (V, w^2)$ be networks of same size. The topological loss $L_{top} (G_1, G_2)$ is defined by the optimal matching cost:
\bq
 L_{top} (G_1, G_2)  = \min_{\tau}  \sum_{b \in I_0 (G_1)}  [  b - \tau (b) ]^2
 +  \sum_{d \in I_1 (G_1)}  [ d - \tau (d) ]^2 
\eq
assuming bijection $\tau:  I_0 (G_1) \cup I_1 (G_1) \rightarrow I_0 (G_2) \cup I_1 (G_2)$ factorizes $\tau = \tau_0 \oplus \tau_1$ into bijections $\tau_0: I_0 (G_1) \rightarrow I_0 (G_1)$ and $\tau_1: I_1(G_1) \rightarrow I_1(G_2)$.
 $L_{top}$ is the modified Wasserstein distance between the barcodes of the two networks.
 It can be shown that the optimal bijection $\tau_0$ is given by matching  the $i$-th smallest birth values in $I_0 (G_1)$ to the $i$-th smallest birth  values in $I_0 (G_2)$ \citep{song.2020.ISBI}. Similarly $\tau_1$ is given by matching the $i$-th smallest death values in $I_1 (G_1)$ to the $i$-th smallest death values in $I_1 (G_2)$. For two networks of different sizes, any unmatched 0D birth or 1D death values in the larger network are matched to the largest or smallest edge weights in the smaller network, respectively.

A topological loss function is given by the Wasserstein distance between the persistence images of two datasets. Other topological loss is also possible. In \citet{bruel2019topology},  topological loss $L_{PD_k}$ on the persistent diagram of $k$-cycle was introduced as
\bq
\label{topological_regularizer}
L_{PD_k} (p, q, i_0) = \sum_{i \geq i_0} (d_i - b_i )^p \left( \frac{d_i + b_i}{2} \right)^q,
\eq
where $(b_i, d_i)$ is the $i$-th barcode and $i_0$ is the index for the $i_0$-th most persistent point. 
The power $p$ can be increased to more strongly penalize the most persistent features, and the power $q$ serves to weight features that are prominent later in the filtration. We can use $L_{PD_k}$ as a regularizer that penalizes on the number of clusters or cycles in a natural way that is difficult to accomplish with more traditional analysis. $L_{PD_k}$ was used in de-noising the number of connected component of MNIST image dataset \citep{bruel2019topology}. Topological regularizers of this type enable us to use the stochastic gradient decent to encourage specific topological features in deep learning tasks.

\subsection{Topological learning}

An important application of deep learning toward brain networks is to uncover patterns in networks, as these patterns can reveal the underlying topological signal.  However, much of the deep learning efforts so far has been limited to using the data structure provided by the {\it local} geometry. For example, the various Euclidean loss functions often used in generative models as well as for regularizations are defined with only  knowledge of the data's local geometry.
In contrast, topology encodes the {\it global} structure of the geometry of data.  The purpose of topological learning toward brain networks is to develop learning frameworks that exploit the topological information of networks explicitly. 

Central to learning tasks such as regression, classification and clustering  is the optimization of a loss function. Given observed data $X_i$ with responses $y_i$, we fit a predictive model with parameters $\widehat \beta$ which will enable us to make a prediction $\widehat{y}_i = f(\widehat{\beta};X_i)$ for each observation. The loss function $L$ assesses the goodness of the fit of the model prediction to the observation, mostly through the Euclidean distance or matrix norms such as the sum of squared residuals 
$$  L (\beta) = \sum_{i=1}^n  \big(  y_i - f(\widehat{\beta};X_i) \big)^2.$$
However, such models are prone to over-fitting if there are more unknown parameters than observations. 
The regularization term or penalty $P(\beta)$ is introduced to avoid such overfitting: 
\bq \label{total_loss}
\widehat{\beta} = {\rm arg~min}_{\beta} L(\beta) + \lambda P(\beta),
\eq
where $\lambda$ is a tuning parameter that controls the contribution of the regularization term. Traditionally common regularizers include $L_1$ and $L_2$ loss functions, however, topological penalties are recently introduced to penalize or favor specific topological features \citep{chen2019topological}. In \citep{song.2020.arXiv}, functional brain networks $G_1, \cdots, G_k$ are regressed  in the model
\bq
\widehat{\Theta} = {\rm arg} \min_{\Theta} \frac{1}{n} \sum_{k=1}^n L_F (\Theta, G_k) +  \lambda L_{top} (\Theta, T),
\eq
where the Frobenius norm $L_F$ measures the goodness-of-fit between the network model $\Theta$ and $P = (V_P , w_P)$ is a network expressing a prior topological knowledge such as the  structural brain network, where functional networks are overlaid. For sufficiently large $\lambda$, all the functional networks will be warped to the structural network $T$ and there will be no topological differences among functional networks.

Topological approaches are relatively unexplored in deep learning. Many past works on incorporating topological features into convolutional neural networks does it implicitly and fail to identify which topological features are learned \citep{clough.2019}. Recently, more explicit topological losses constructed from persistent homology have seen more success in explicitly learning underlying topology of data. Applications of topology to deep learning include the extraction of topological features for better learning \citep{pun2018persistent, Cole:2020hjx}, using differentiable properties of persistence to define topological loss \citep{clough.2019,Edelsbrunner2010,song.2020.ISBI} and persistence images \citep{Cole:2020hjx}, deep learning interpretability \citep{gabrielsson2019exposition,gabella2019topology}.

In other applications, we can incorporate {\em topological priors} in the model or training data to improve the performance in learning tasks \citep{chen2019topological,bruel2019topology}. In image segmentation, standard loss functions is often constructed at the voxel level ignoring the higher-level structures \citep{clough.2019}. However, it may be necessary to enforce global topological prior to avoid the additional image processing to correct  topological defects in the segmentation results \citep{chung.2015.MIA}.

 Topology is also used in designing more robust deep learning models \citet{bruel2019topology}. Adversarial attacks in machine learning try to deceive models by supplying deceptive input or outlying data \citep{huang.2017}. Machine learning techniques mainly assume the the training and test data are generated from the statistical distributions. By designing the input data that violates such statistical assumptions, we can fool the learning model. Neutral networks are vulnerable to adversarial attacks which may be imperceptible to the human eye, but can lead the model to misclassify the output \citep{chakraborty2018adversarial}. It has been speculated  that models based on topological features are more robust against adversarial attacks \citep{gabrielsson2019exposition,carlsson2018topological}. A topological layer can also be placed in neural networks to prevent from adversarial attacks and enhance topological fidelity. In   \citet{bruel2019topology}, it was shown that the success rate of adversarial attacks is much smaller on networks with a topological layer. Even thought  the concept of adversarial attacks have not been popular in brain network modeling,  building deep learning models on brain networks that are robust against adversarial attacks would be highly useful.

\section{Conclusion \& Disucssion}

\begin{figure}[t]
\begin{center}
\includegraphics[width=1\linewidth]{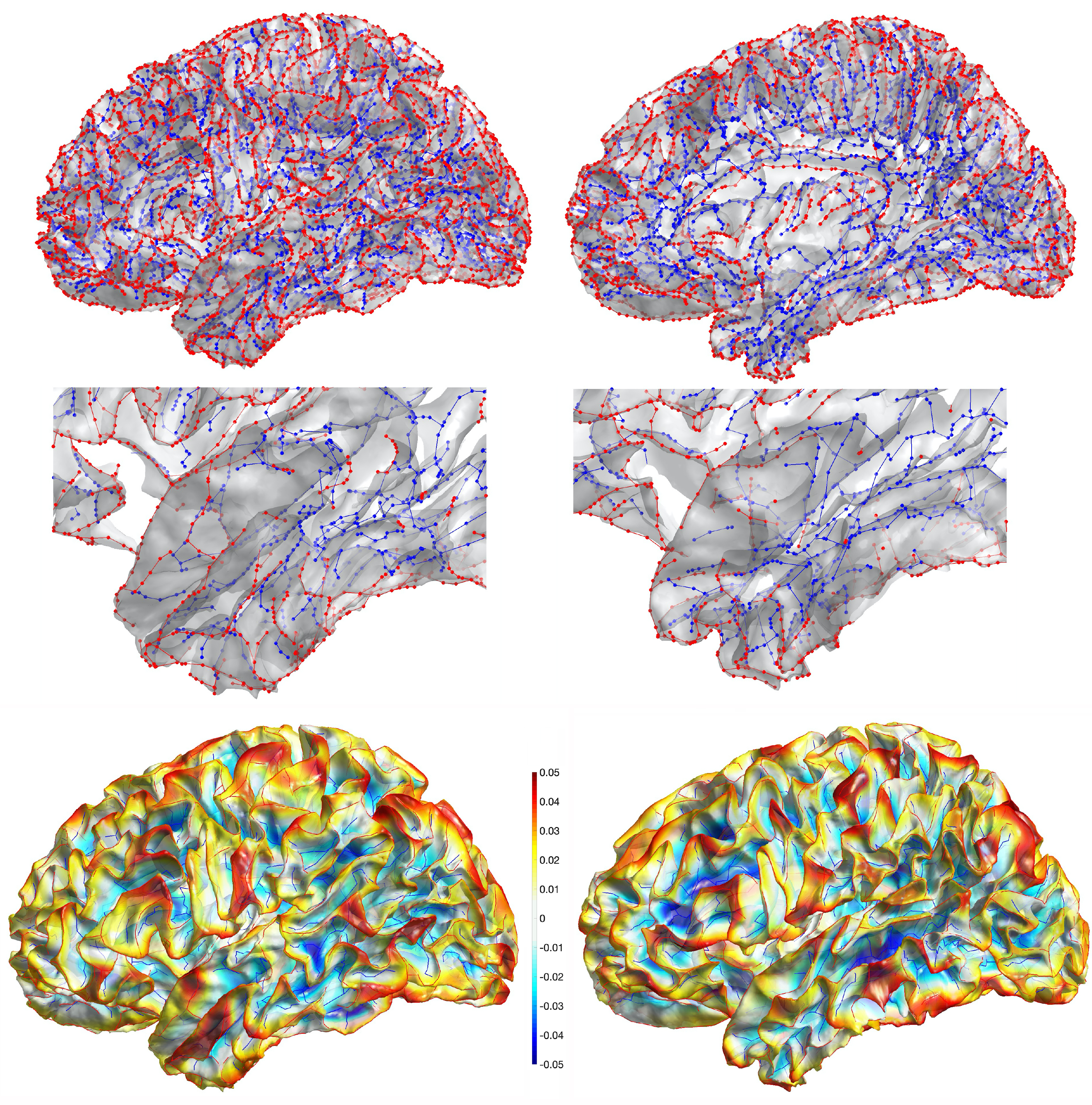}
\caption{Top: Sulcal (blue) and gyral (red) pattern of  two different subjects.   The sucal pattern can be represented as a forest, a collection of trees  \citep{huang.2020.TMI}. Since the nodes of the forests are the surface mesh vertices, the forest can be represented as about 300000 $\times$ 300000 sparse adjacency matrices. Computing similarity measures between the sulcal forest is not trivial and may require topological distances. Middle: enlargement of the temporal lobe. Bottom: The weighted spherical harmonic representation with degree 70 and heat kernel bandwidth $0.001$ \citep{chung.2008.TMI} smoothing the node value  where sulci are assigned value -1 and gyri are assigned value 1. The smoothing obliviates the need for matching heterogeneous data structure and comparisons of the different sulcal pattern across subject can be done at the mesh vertex level. Such local geometric solution is not readily available for brain networks. 
}
\label{fig:sulcaltree}
\end{center}
\end{figure}

In this review paper, we surveyed various topological distance and loss functions based on persistent homology. 
We showed how such distance and loss functions can be used in wide variety of learning applications including regression, clustering and classification. ManyTDA approaches can successfully differentiate the topological network differences. However, often it is unclear where the difference is localized within the networks. In diagnostics related problems such as determining if a subject belong to particular clinical population, it is important to determine the topological differences; however, more important questions is localizing the source of differences. Since there is no one-to-one correspondence between the topological features and the original data, it is often not possible to localize the signals, which has been the biggest limitation of the TDA methods in brain imaging applications. The future development TDA should be toward this important but very difficult question. 

\begin{figure*}[t]
\begin{center}
\includegraphics[width=1\linewidth]{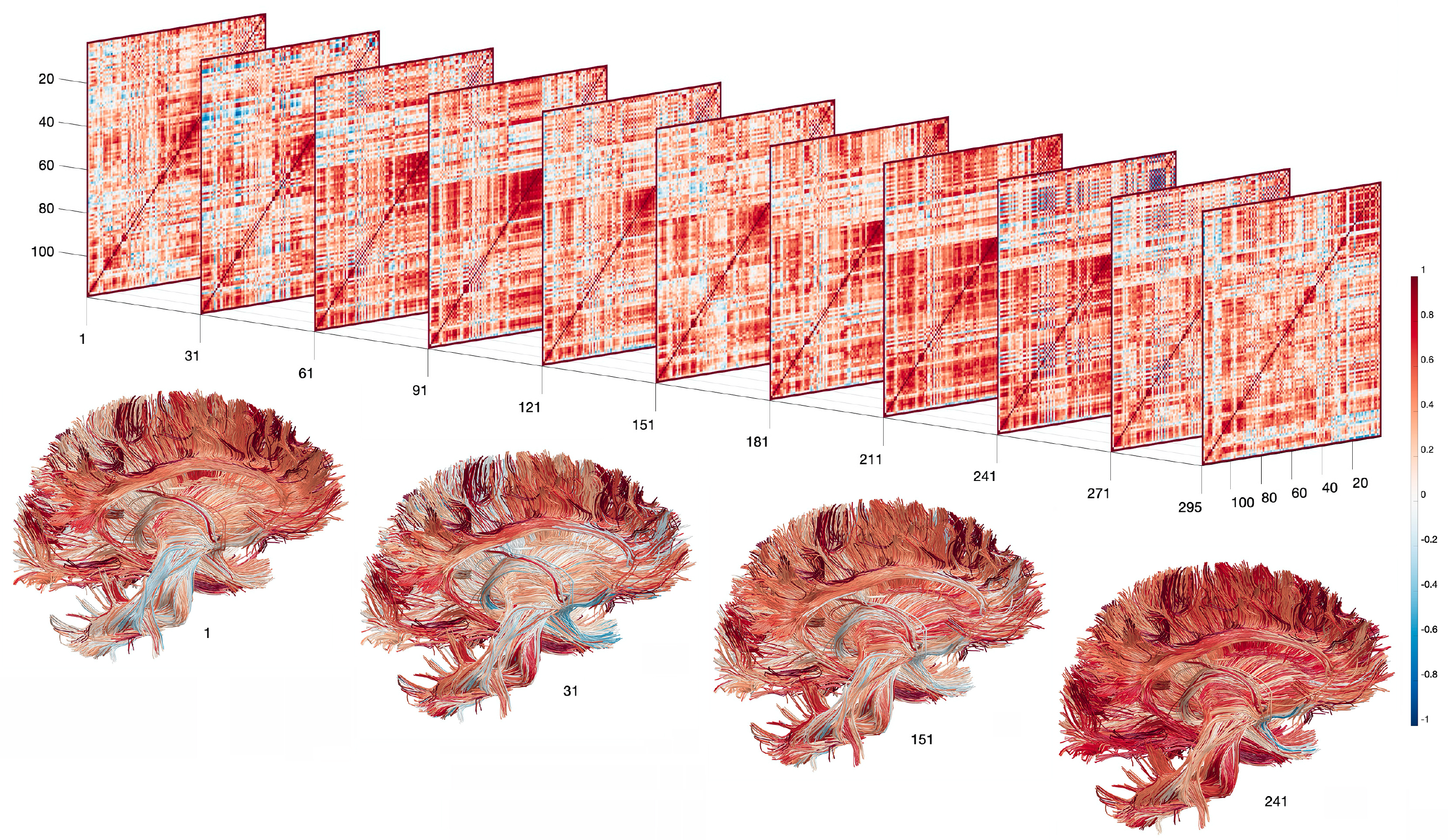}
\caption{Top: Dynamically changing correlation matrices computed from rs-fMRI using the sliding window of size 60 for a subject \citep{huang.2019.DSW}. Bottom: The constructed correlation matrices are superimposed on top of the white matter fibers, which are colored based on correlations between parcellations. The structural brain network is a more or less a  tree like structure without many loops or cycles while the functional bran networks can have many loops and cycles. Thus, aligning functional brain networks on top of structural brain network directly will not work. It requires a new topological approach for aligning topologically different networks.}
\label{fig:dynamicTDA}
\end{center}
\end{figure*}

Compared to TDA methods, geometric methods are more adapt at detecting localized signals in brain imaging. Figure \ref{fig:sulcaltree} displays the sucal and gyral trees obtained from the brain surface mesh \citep{huang.2019.MICCAI}, where trees are teated as heat source with value +1 on gyral trees and heat sink with value -1 on sulcal trees. Then Isotropic diffusion is applied to produce the smooth map of sulcal and gyral trees. Such smooth maps can be easily compared across different subjects. For instance, two-sample $t$-statistic at each mesh vertex is performed localizing group differences. Such localized signal detection is difficult if not impossible with many existing TDA methods. 

Persistent homology features are by definition global summary measures and they might be more useful for tasks that do not involves identifying the source of signal differences. It might be more useful in discrete decision making tasks such as clustering and classifications. In fact TDA has begin to be more useful in deep learning \citep{chen2019topological}.

The paper reviewed how TDA can be used in statistic brain networks that do not change over time. For dynamically changing brain networks over time, it is unclear how TDA can be applied yet. Consider the dynamically changing correlating matrices
of resting-state fMRI (Figure \ref{fig:dynamicTDA}). At each each correlation matrix, we can apply persistent homology and obtain PD, which is the discrete representation of connectivity at one particular time point. It is unclear how to integrate such discrete topological representation over multiple time points in a continuous fashion. Dynamic-TDA as introduced in \citet{song.2020.ISBI}  encodes rsfMRI as a time-ordered sequence of Rips complexes and their corresponding barcodes in studying dynamically changing topological patterns over time. Dyanmic-TDA may provide some suggestion for extending TDA to time varying brain networks. This is left as promising future research direction.

\section*{Acknowledgements}
This study was supported by NIH grants R01 EB022856 and R01 EB028753 and NSF grant MDS-2010778. We would like to thank Andery Gritsenko of Northeastern University for the computation of the largest cycle in a graph. We also like to thank Alex Cole of University of Amsterdam for the discussion of the witness complex.  We thank Hyekyung Lee of Seoul National University Hospital for discussion on sparse filtrations and kernel distances.  We would like to thank Robin Henderson of Newcastle University, U.K. for providing the coordinates and connectivity information of nodes of the binary tree used in \citet{garside.2020}. We would like to thank Hill Goldsmith of University of Wisconsin for providing one subject rsfMRI data used in the figure, which came from the study \citet{huang.2019.DSW}. We thank Taniguchi Masanobu of Waseda University for discussion on canonical correlations. 

\section*{\refname}

\bibliographystyle{elsarticle-harv} 
\bibliography{reference.2021.01.21,reference_shiu_2021,ADS_ref,bib_tda_survey,reference.2020.09.15cole.bib}

\end{document}